\documentclass[5p]{elsarticle}
\usepackage{amssymb}
\usepackage{amsmath}
\usepackage[figuresright]{rotating}
\usepackage{color}

\begin{document}

\begin{frontmatter}

\title{{\bf\boldmath The polarization observables 
$T$, $P$, and $H$ and their impact on $\gamma p\to p\pi^0$
multipoles}}

%1 = HISKP
%2 = PI
%3 = GATCHINA
%4 = BOCHUM
%5 = GIESSEN
%6 = FSU
%7 = BASEL

\author[label1]{J.~Hartmann}
\author[label2]{H.~Dutz}
\author[label1,label3]{A.V.~Anisovich}
%%\author[label2]{B.~Bantes}
\author[label1,label3]{D.~Bayadilov}
\author[label1]{R.~Beck}
\author[label1]{M.~Becker}
\author[label3]{Y.~Beloglazov}
\author[label4]{A.~Berlin}
\author[label4]{M.~Bichow}
\author[label1]{S.~B\"ose}
\author[label5,label1]{K.-Th.~Brinkmann}
%\author[label7]{Th.~Challand}
\author[label6]{V.~Crede}
\author[label7]{M.~Dieterle}
%\author[label5]{F.~Dietz}
\author[label2]{H.~Eberhardt}
\author[label2]{D.~Elsner}
%\author[label2]{R.~Ewald}
\author[label2]{K.~Fornet-Ponse}
\author[label5]{St.~Friedrich}
\author[label2]{F.~Frommberger}
\author[label1]{Ch.~Funke}
\author[label1]{M.~Gottschall}
\author[label3]{A.~Gridnev}
\author[label1]{M.~Gr\"uner}
\author[label2]{St.~G\"ortz}
\author[label5,label1]{E.~Gutz}
\author[label1]{Ch.~Hammann}
\author[label2]{J.~Hannappel}
\author[label5]{V.~Hannen}
\author[label4]{J.~Herick}
\author[label2]{W.~Hillert}
\author[label1]{Ph.~Hoffmeister}
\author[label1]{Ch.~Honisch}
%\author[label7]{\color{red}I.~Jaegle}
\author[label2]{O.~Jahn}
\author[label2]{T.~Jude}
%\author[label1]{C.~Jost}
\author[label7]{A.~K\"aser}
\author[label1]{D.~Kaiser}
\author[label1]{H.~Kalinowsky}
\author[label1]{F.~Kalischewski}
%\author[label2]{S.~Kammer}
\author[label1]{P.~Klassen}
\author[label7]{I.~Keshelashvili}
\author[label2]{F.~Klein}
\author[label1]{E.~Klempt}
\author[label1]{K.~Koop}
\author[label7]{B.~Krusche}
\author[label1]{M.~Kube}
\author[label1]{M.~Lang}
\author[label3]{I.~Lopatin}
%\author[label7]{Y.~Maghrbi}
\author[label5]{K.~Makonyi}
\author[label2]{F.~Messi}
\author[label5]{V.~Metag}
\author[label4]{W.~Meyer}
\author[label1]{J.~M\"uller}
\author[label5]{M.~Nanova}
\author[label1,label3]{V.~Nikonov}
\author[label3]{D.~Novinski}
\author[label5]{R.~Novotny}
\author[label1]{D.~Piontek}
\author[label2]{S.~Reeve}
\author[label1]{Ch.~Rosenbaum}
\author[label4]{B.~Roth}
\author[label4]{G.~Reicherz}
\author[label7]{T.~Rostomyan}
\author[label2]{St.~Runkel}
\author[label1,label3]{A.~Sarantsev}
\author[label1]{Ch.~Schmidt}
\author[label2]{H.~Schmieden}
\author[label1]{R.~Schmitz}
\author[label1]{T.~Seifen}
\author[label1]{V.~Sokhoyan}
\author[label1]{Ph.~Th\"amer}
\author[label1]{A.~Thiel}
\author[label1]{U.~Thoma}
\author[label1]{M.~Urban}
\author[label1]{H.~van~Pee}
\author[label1]{D.~Walther}
\author[label1]{Ch.~Wendel}
\author[label4]{U.~Wiedner}
\author[label1,label6]{A.~Wilson}
\author[label1]{A.~Winnebeck}
\author[label7]{L.~Witthauer}
\author{\\[2ex]CBELSA/TAPS Collaboration}

\address[label1]{Helmholtz--Institut f\"ur Strahlen-- und Kernphysik, Universit\"at Bonn, 53115 Bonn, Germany}
\address[label2]{Physikalisches Institut, Universit\"at Bonn, 53115 Bonn, Germany}
\address[label3]{Petersburg Nuclear Physics Institute, Gatchina, 188300 Russia}
\address[label4]{Institut f\"ur Experimentalphysik I, Ruhr--Universit\"at Bochum, 44780  Bochum, Germany}
\address[label5]{II.~Physikalisches Institut, Universit\"at Gie{\ss}en, 35392 Gie{\ss}en, Germany}
\address[label6]{Department of Physics, Florida State University, Tallahassee, Florida 32306, USA}
\address[label7]{Department Physik, Universit\"at Basel, 4056 Basel, Switzerland}

\begin{abstract}
Data on the polarization observables $T$, $P$, and $H$ for the
reaction $\gamma p\to p\pi^0$ are reported. Compared to earlier data 
from other experiments, our data are more precise 
and extend the covered range in energy and angle substantially. The
results were extracted from azimuthal asymmetries measured using a
transversely polarized target and linearly polarized photons.
The data were taken at the Bonn electron stretcher accelerator ELSA
with the CBELSA/TAPS detector. Within the Bonn-Gatchina partial wave
analysis, the new polarization data lead to a significant narrowing 
of the error band for the multipoles for neutral-pion photoproduction.
\vspace{2mm}
%\begin{keywords}
%\end{keywords}
%----------end of abstract
\end{abstract}


\end{frontmatter}

%%
%% Start line numbering here if you want
%%
% \linenumbers

%% main text
\section{Introduction}
%
%{\color{red} Some general introduction missing .... presently direct start 
%with pol. obs. .......... }\\[+2ex]
%
The measurement of the two double polarization observables $G$
\cite{Thiel:2012yj} and $E$  \cite{Gottschall:2014} in
photoproduction of neutral pions revealed significant differences 
between the data and the predictions from analyses such as MAID~\cite{Drechsel:2007if}, 
SAID~\cite{Workman:2011vb}, and BnGa~\cite{Anisovich:2011fc}. 
Partly, large discrepancies were observed even
at rather low photon energies. This was surprising since the
reaction $\gamma p\to p\pi^0$ is certainly the best studied
photoproduction process. These discrepancies underline the
importance of polarization observables for an interpretation
of photoproduction data.

In this letter, we report a measurement of further polarization
observables, called $T$, $P$, and $H$, for the reaction
 \begin{equation}
 \gamma p\to p\pi^0\,. 
 \label{reac}
 \end{equation}
All three observables were determined simultaneously from the 
same measurement and 
provide the next important step toward a better understanding 
of $\pi^0$ photoproduction. 
The target asymmetry $T$ is a measure of the azimuthal asymmetry when
the target nucleon carries polarization $p_{\rm T}$ in a direction
perpendicular to the beam axis. $P$, often termed the recoil polarization 
observable, is a measure of the induced polarization of the recoiling
nucleon. % in a direction perpendicular to the reaction plane. 
Here, $P$ is determined from a double polarization measurement rather than from
an experimentally more challenging direct measurement of the recoil 
polarization. 
This has the advantage that $P$ can be determined in the very same 
measurement for almost the full solid angle, rather than by measuring 
$P$ for specific points in angle and energy, as it has been done in the past. 
The observables $P$ and $H$ can be determined from azimuthal asymmetries
using measurements with linearly polarized photons and transversely polarized 
target nucleons having polarization $p_{\rm T}$.
Part of the data presented here were used as a basis for an energy independent
partial wave analysis (PWA)~\cite{jan_prl}. They are now included in the BnGa PWA
and multipoles for reaction~(\ref{reac}) were determined. The multipoles were 
compared to those from MAID, SAID, Juelich2015~\cite{Ronchen:2015} 
and earlier BnGa PWA-solutions.

\section{The experiment}

The experiment was performed at the Bonn Electron Stretcher
Accelerator ELSA \cite{Hillert:2006yb}. Linearly polarized photons
were produced by scattering a 3.2\,GeV electron beam off a
diamond crystal \cite{Elsner:2008sn}. The crystal was oriented to
position the coherent edge at 950\,MeV, leading to a polarization maximum of
$p_{\gamma}=65$\% at 850\,MeV which declined to 40\% at
700\,MeV. Two perpendicular settings of the beam polarization plane were
used (named $\parallel$ and $\perp$). Photon energies were
measured in a tagging system described in Ref.~\cite{Elsner:2008sn}.

The photon beam hits a butanol (${\rm C_4H_{9}OH}$) target with
transversely polarized protons \cite{Dutz:1999} with a mean
proton polarization of $p_{\rm T}\approx 75$\%. Data were taken with
two opposite settings of the target polarization direction (named
$\uparrow$ and $\downarrow$). 
%The butanol target was replaced by a
%foam carbon target to determine the contributions from the
%unpolarized carbon and oxygen nuclei in butanol.

\begin{figure}[t]
\begin{center}
\includegraphics[width=\linewidth]{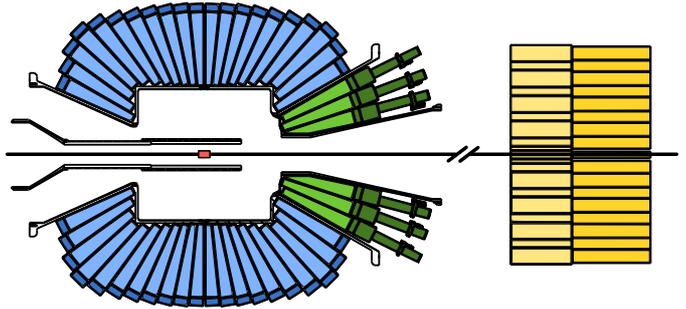}
\end{center}
\caption{\label{pic:detector}(Color online) The central part of the
detector system. The 1320 CsI(Tl) crystals (blue and green) are read
out via wavelength shifters and photodiodes (blue) or via 
photomultipliers (green), the 216 BaF$_2$ crystals (yellow) in
forward direction are read out with photomultipliers. The target is surrounded by
a 3-layer-scintillating fiber detector. To detect charged particles,  
scintillators are also placed in front of the forward CsI(Tl) (green) and 
TAPS crystals (yellow).}
\end{figure}

The incoming photons may produce a $\pi^0$ in the reaction~(\ref{reac}). 
The neutral pions were reconstructed from their
$\pi^0\to 2\gamma$ decays in the Crystal Barrel (1320 CsI(Tl)-crystals)~\cite{Aker:1992ny}
and TAPS (216 BaF$_2$ crystals)~\cite{Novotny:1991ht,Gabler:1994ay} electromagnetic
calorimeters (see Fig.~\ref{pic:detector}) which cover almost the full
angular range down to $\theta = 1^\circ$ in the forward direction. Protons from reaction~(\ref{reac})
were detected in a three-layer cylindrical scintillation detector
with 513 fibers \cite{Suft:2005cq} surrounding the target, in 180 small
organic scintillators in front of 90 forward CsI(Tl) crystals 
covering the angular range from 27.5$^\circ$ to 11.2$^\circ$, and in 
organic scintillators mounted in front of each of the BaF$_2$ crystals. A CO$_2$ Cherenkov detector was
installed in front of the BaF$_2$ crystals to identify background from electromagnetic reactions.
The first-level trigger was derived from the tagger, the fiber detector, the forward calorimeters,
and from the CO$_2$ Cherenkov detector as a veto; a second-level trigger used a FAst Cluster Encoder
(FACE) \cite{Flemming:2001} and selected events with at least two distinct calorimeter
hits in the full detector assembly. 
%Details of the detector
%performance and of the analysis chain can be found elsewhere
%\cite{vanPee:2007tw}.

\begin{figure}[!t]
\begin{center}
\includegraphics[width=\linewidth]{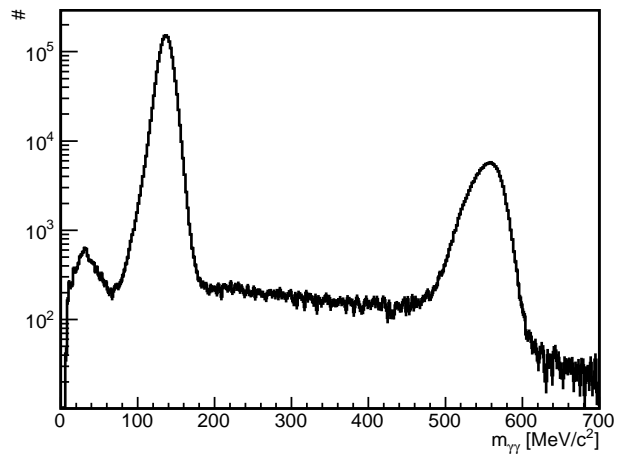}
\end{center}
\vspace*{-3mm}
\caption{\label{pic:mgg} The invariant $\gamma\gamma$ mass distribution for the butanol data, after all cuts discussed (see text) have been applied. The final data sample of $p\pi^0$ events has a background contamination of less than 1\%.}
\end{figure}

\begin{figure*}[t]
\begin{center}
\begin{tabular}{ccc}
\includegraphics[width=0.33\linewidth]{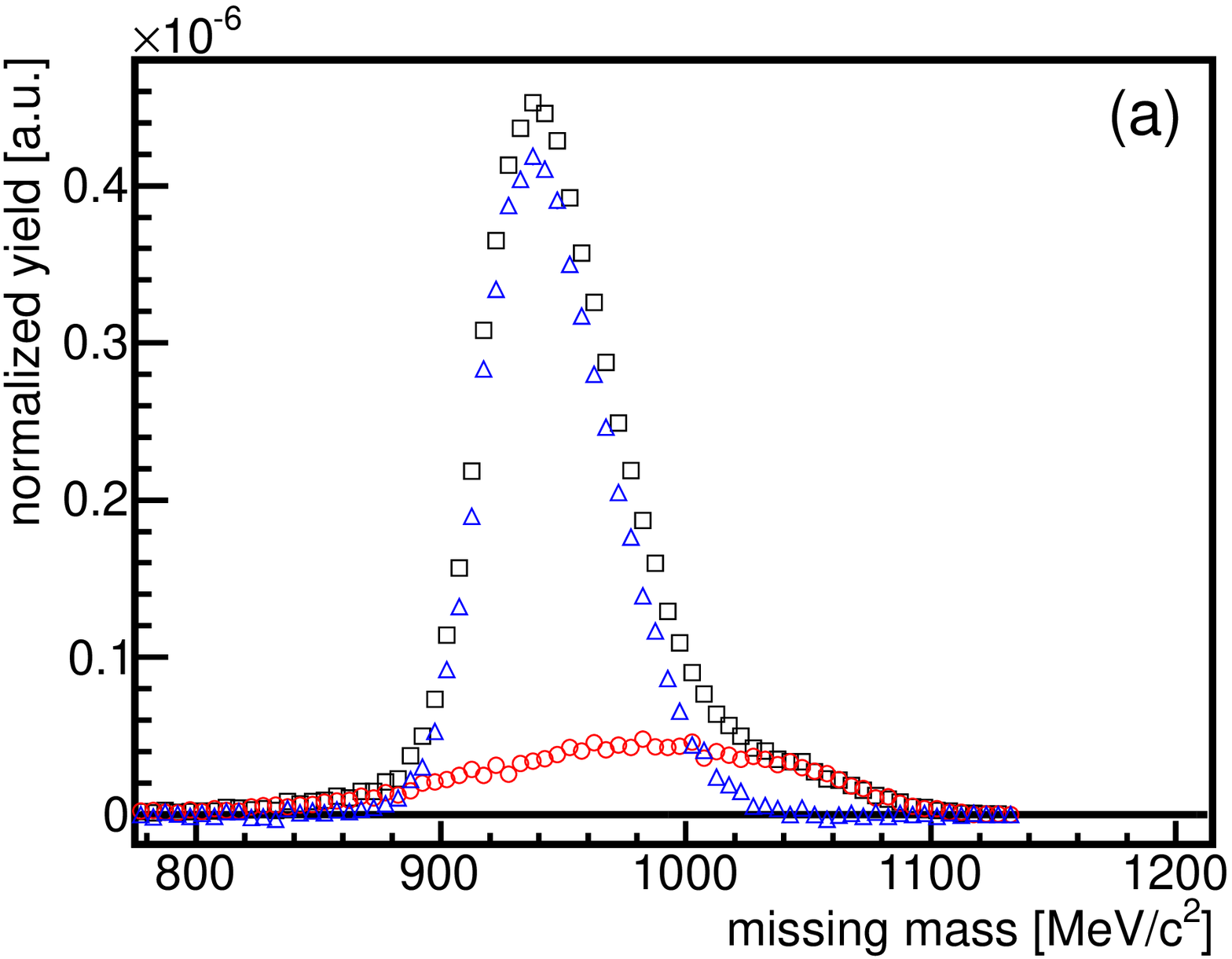} &
\hspace{-0.02\linewidth}\includegraphics[width=0.33\linewidth]{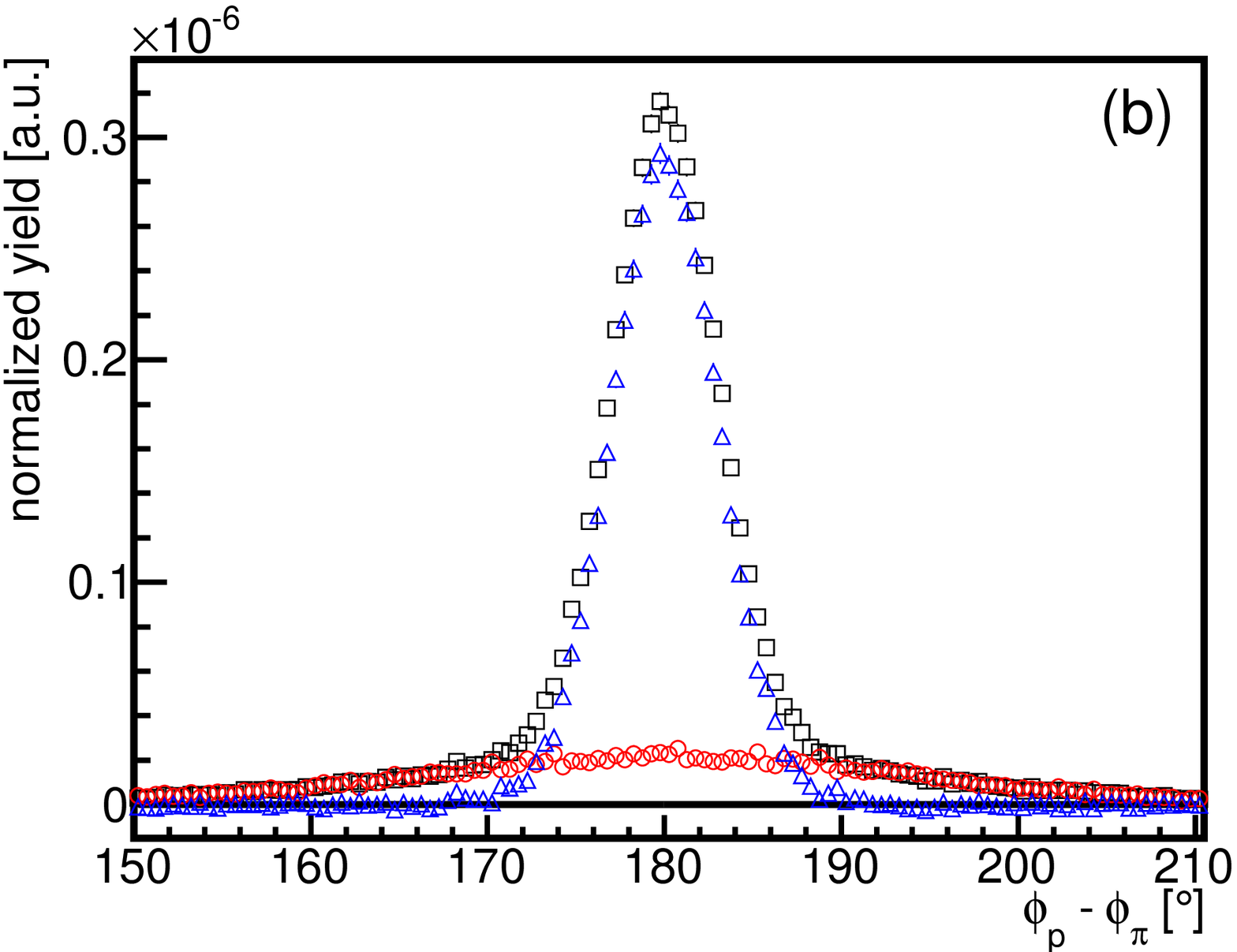} &
\hspace{-0.02\linewidth}\includegraphics[width=0.33\linewidth]{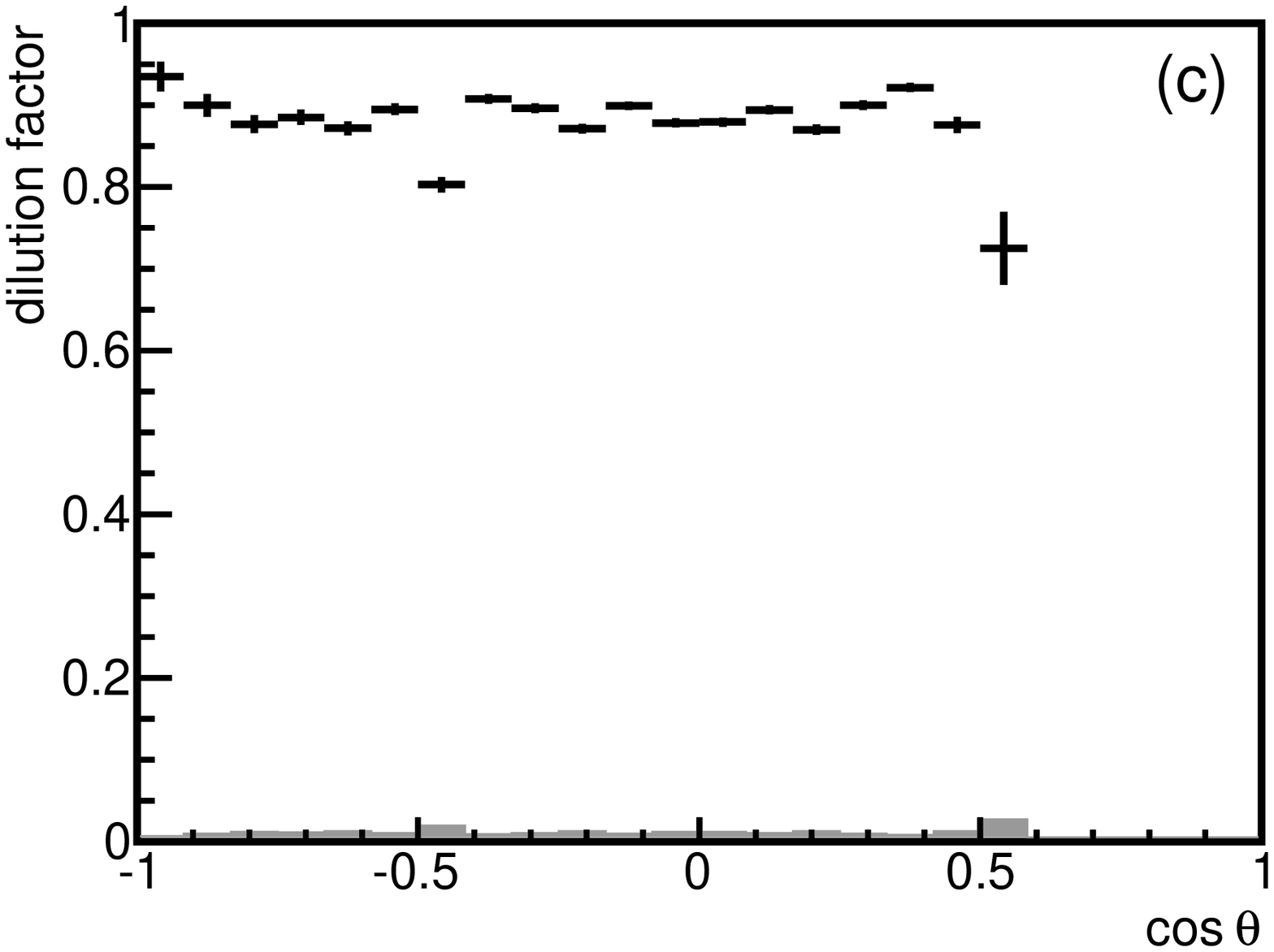} \\
\includegraphics[width=0.33\linewidth]{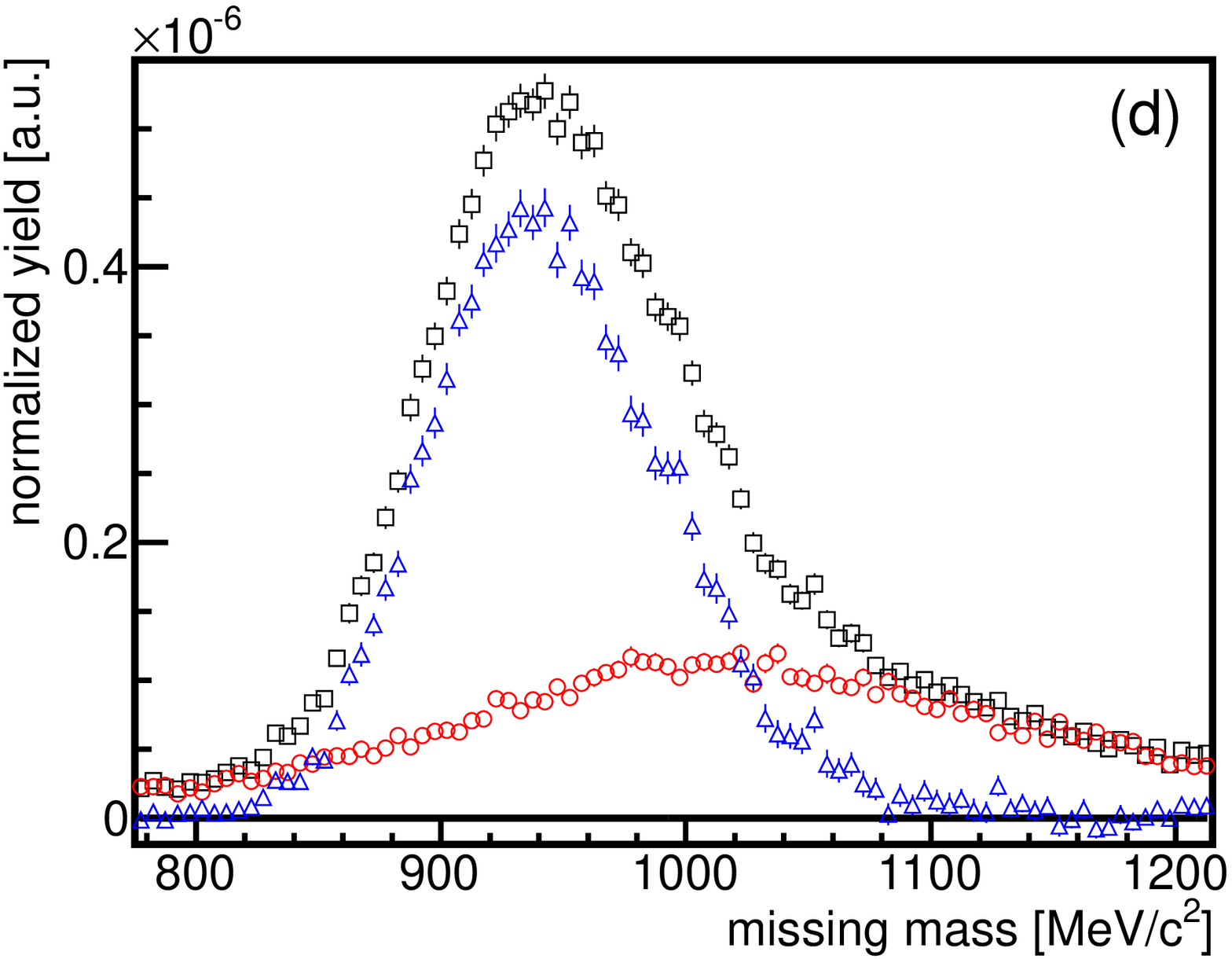} &
\hspace{-0.02\linewidth}\includegraphics[width=0.33\linewidth]{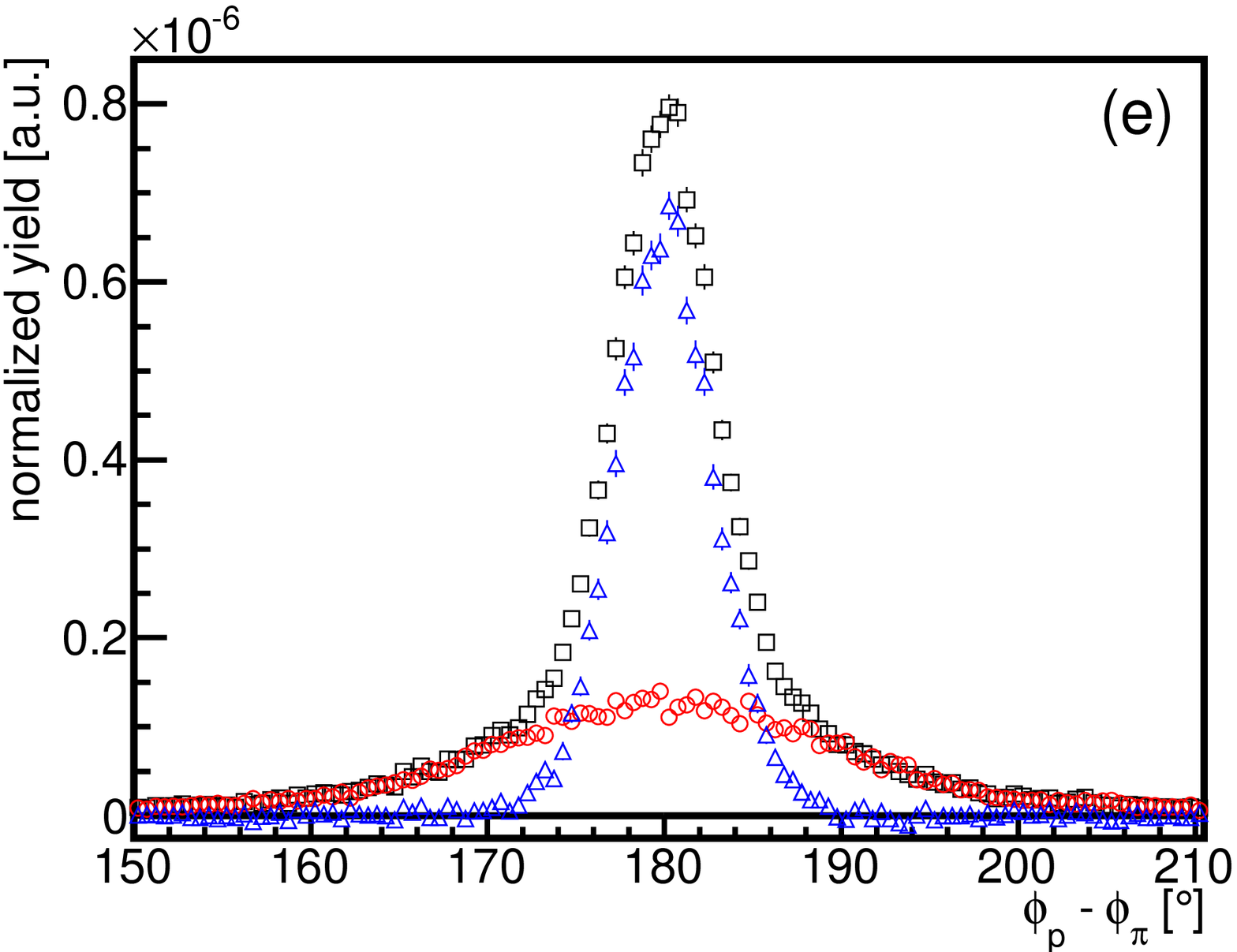} &
\hspace{-0.02\linewidth}\includegraphics[width=0.33\linewidth]{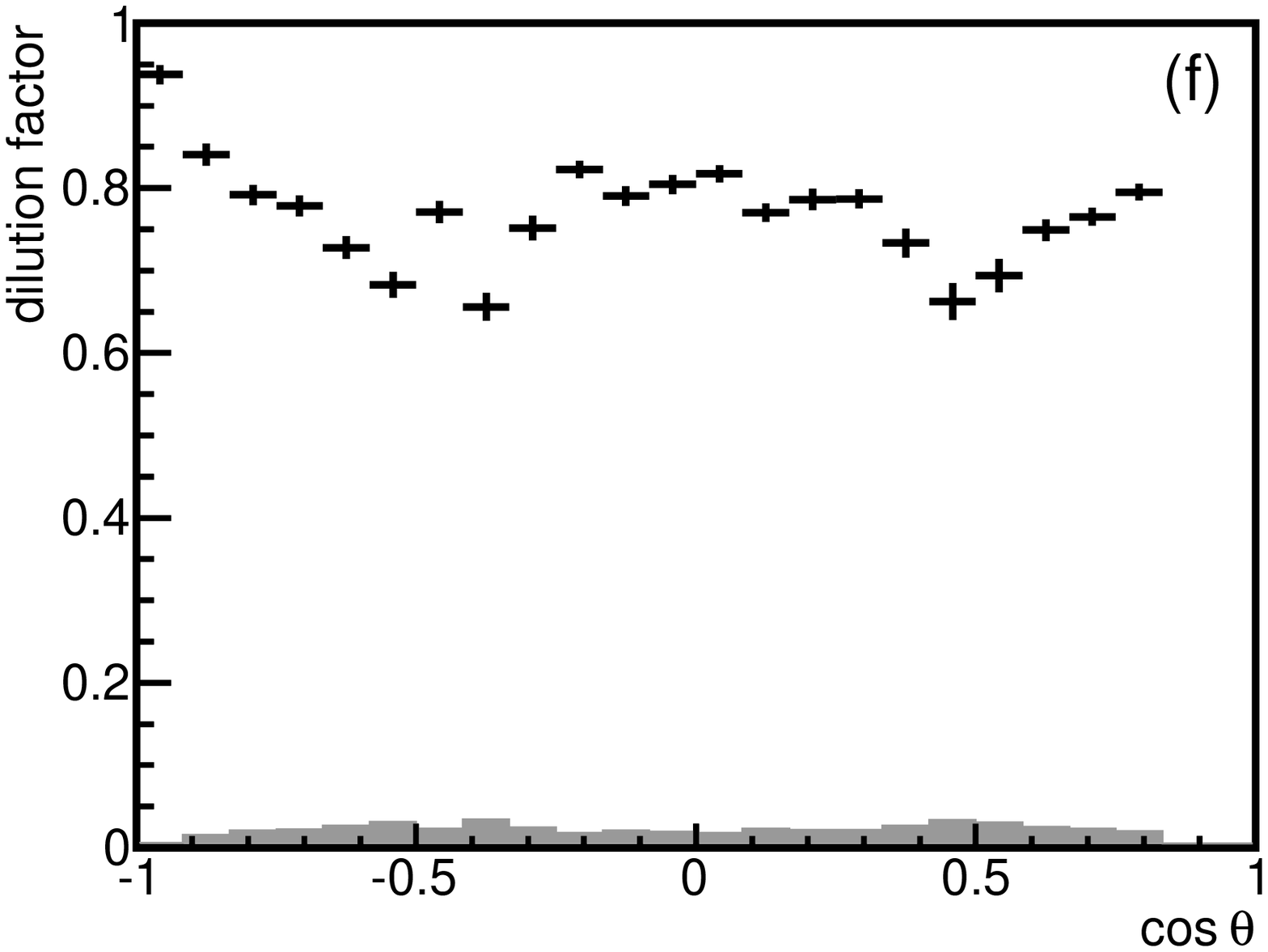}
\end{tabular}
\end{center}
\caption{\label{pic:dilution}(Color online) (a,d) The missing mass
distribution, with the proton as the missing particle,
and (b,e) the azimuthal angle difference of $\pi^0$ and proton for
reaction (1), for a $\gamma p$ invariant mass of $W = 1.46$\,--\,1.48 
(top) and 1.82\,--\,1.94\,GeV (bottom); butanol
({\scriptsize$\square$}), scaled carbon ({\color{red}$\circ$}), and the
difference ({\color{blue}$\triangle$}). The distributions are shown 
after all other cuts discussed in the text are applied. From these distributions,
the dilution factor (c,f) is determined. The gray band indicates 
the systematic uncertainty in the dilution factor  due to uncertainties
in the determination of the carbon scaling factor $s$. 
Since only events with all three particles detected in the calorimeter are 
considered an acceptance hole occurs for large cos($\theta$). 
The observed structures in the dilution factor are due to a combined
effect of reduced efficiencies for clusters impinging onto detector
boundaries and Fermi smearing.}
\end{figure*}

For further analysis only events with three distinct calorimeter hits were used. Adding events where only the two photons were measured would lead to an increased background contribution from other channels since several of the cuts discussed below can no longer be applied. For the events with three calorimeter hits, all three possible combinations were treated as $p\gamma\gamma$ candidates, with the proton being treated as a missing particle.
Kinematic cuts were applied to ensure momentum conservation. Examples for the missing mass and 
azimuthal angle difference distributions are shown in Fig.~\ref{pic:dilution}. 
Energy- and angle-dependent $\pm2\sigma$ cuts were applied based on the corresponding distributions.
In addition, a $\pm2\sigma$-cut on the polar-angle difference between the 
directions of the missing proton and the measured proton candidate was performed, and 
${\gamma}{p} \to p\pi^0$ events were selected by a $\pm2\sigma$ cut 
on the invariant $\gamma\gamma$ mass. Finally, a time coincidence was 
required between the tagger hit and the reaction products and a random-time 
background subtraction was performed. This resulted in a final data 
sample containing a total of $1.4$ million $p\pi^0$ events. The background contamination
was estimated from the invariant $\gamma\gamma$ mass spectrum (see Fig.~\ref{pic:mgg}), assuming a linear behaviour of the background under the peak. It is less than 1\% for all energies and angles.
The selected events for each of the four combinations of beam and 
target polarization directions were
normalized to the corresponding photon flux and 
polarization degree for further analysis. 

Since a butanol target was used, not only polarized and unpolarized
free protons contributed to the count rates but also reactions
occurring on the bound unpolarized nucleons of the carbon and oxygen
nuclei. Additional measurements using a carbon foam target were
performed to determine the so-called dilution factor
$d(W,\cos{\theta})$ 
\begin{align}
\label{dil_factor}
d(W,\cos{\theta}) &= \frac{N_\text{free}(W,\cos{\theta})}{N_\text{butanol}(W,
\cos{\theta})} \nonumber\\
	&= \frac{N_\text{butanol}(W,\cos{\theta})-N_\text{bound}(W,\cos{\theta})}
{N_\text{butanol}(W,\cos{\theta})}
\end{align}%
which assumes that the nucleons bound in carbon and
oxygen show the same response to the impinging photons. 
The carbon foam target had the same size as the butanol target and
approximately the same target area density as the carbon and oxygen
part in the butanol. The carbon target replaced the butanol target in the frozen spin
cryostat to match the experimental conditions of the butanol measurement
as closely as possible. 
%
%The flux-normalized carbon yield was then
%compared to the flux-normalized butanol data using the coplanarity
%distributions. 
%The fluxes were normalized in its wings $|\Delta\phi
%-180^\circ|> 20^\circ$ where no contributions from free protons can
%be expected. Using the global normalization factor $s$ as
%$N_{bound}(W,\cos{\theta}) = s \cdot N_{carbon}(W,\cos{\theta})$,
%the dilution factor (\ref{dil_factor}) can be calculated. Local
%(energy and angle dependent) deviations from the global scaling
%factor contributed to the final systematic uncertainty.
%
The flux-normalized carbon yield was compared to the flux-normalized 
butanol data using the distribution of the angle between the $\pi^0$ 
and proton in the azimuthal plane
%the $\pi^0$ and proton azimuthal angle difference distribution 
outside the region where
contributions from free protons can be expected (see Fig.~\ref{pic:dilution}). 
Counting the yields for $|\Delta\phi -180^\circ|> 20^\circ$, a global scaling factor 
$s = 1.13 \pm 0.01_\text{stat.} \pm 0.11_\text{sys.}$ was determined based on which the dilution factor (Eq.~\ref{dil_factor}) 
was calculated using 
\begin{equation}
N_\text{bound}(W,\cos{\theta}) = s \cdot N_\text{carbon}(W,\cos{\theta}). 
\end{equation}
Taking the slightly different densities of the butanol and carbon targets into account,
one would expect a scaling factor of $s \approx 1.1$, which is in agreement
with the value obtained from the data. Energy- and angle-dependent deviations from the
global scaling factor $s$ were investigated by determining $s(W,\cos\theta)$ indepently
for each bin and comparing it to the global value. The observed deviations are of the
same magnitude as their statistical uncertainty. The global value is used, with the
deviation contributing to the systematic uncertainty.

Figure~\ref{pic:dilution} shows, for two energy bins, the
azimuthal angle difference and the missing mass distributions of the butanol and scaled
carbon data as well as their resulting difference. From these
distributions the dilution factor as function of $\cos\theta$ for 
each energy bin was determined. The dilution factor
is quite large, around 0.9 at low energies and decreasing to around 0.6
at higher energies. Note that a dilution factor $d=1$ corresponds to 
non-existent carbon background and therefore to no dilution. 
The observed structures (Fig.~\ref{pic:dilution}) are due to a combined
effect of reduced efficiencies for clusters impinging onto detector
boundaries and Fermi smearing. They are reproduced in Monte Carlo simulations. 
At higher energies, the reduced missing mass
resolution required a wider cut. Therefore the carbon contribution remaining
after all cuts increased significantly, resulting in a smaller dilution factor.

\section{The polarization observables}

In the coordinate frame of the detector system, with $\alpha$ being the azimuthal angle of the
photon beam polarization plane in the $\parallel$ setting, $\beta$
the azimuthal angle of the target polarization vector in the
$\uparrow$ setting, and $\phi$ the azimuthal angle of the produced $\pi^0$,
the differential cross section can be written as 
\begin{align}
\label{pol-h}
\frac{d\sigma}{d\Omega} = & \left(\frac{d\sigma}{d\Omega}\right)_0 \cdot \{1 - p_{\gamma}\Sigma\cos(2(\alpha-\phi)) + p_{\rm T} T \sin(\beta-\phi) \nonumber\\
  & - p_{\gamma} p_{\rm T} P\cos(2(\alpha-\phi))\sin(\beta-\phi) \nonumber\\
  & + p_{\gamma} p_{\rm T} H\sin(2(\alpha-\phi))\cos(\beta-\phi) \}.
\end{align}
%using the sign convention of {\color{red} BnGa (welche Quelle?)}

\begin{figure*}[t]
\begin{tabular}{ccc}
\hspace{-2mm}\includegraphics[width=0.35\linewidth]{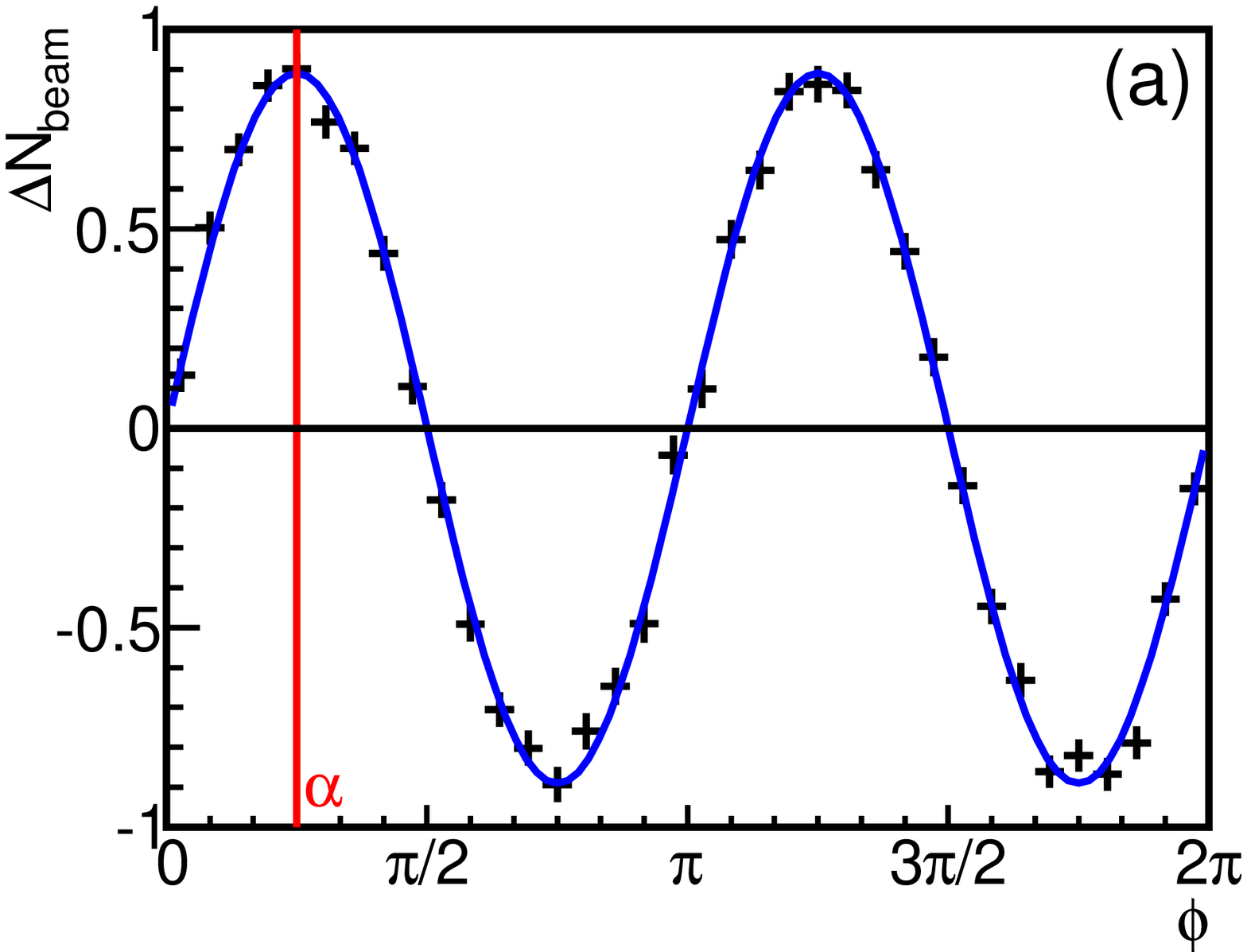}&
\hspace{-7mm}\includegraphics[width=0.35\linewidth]{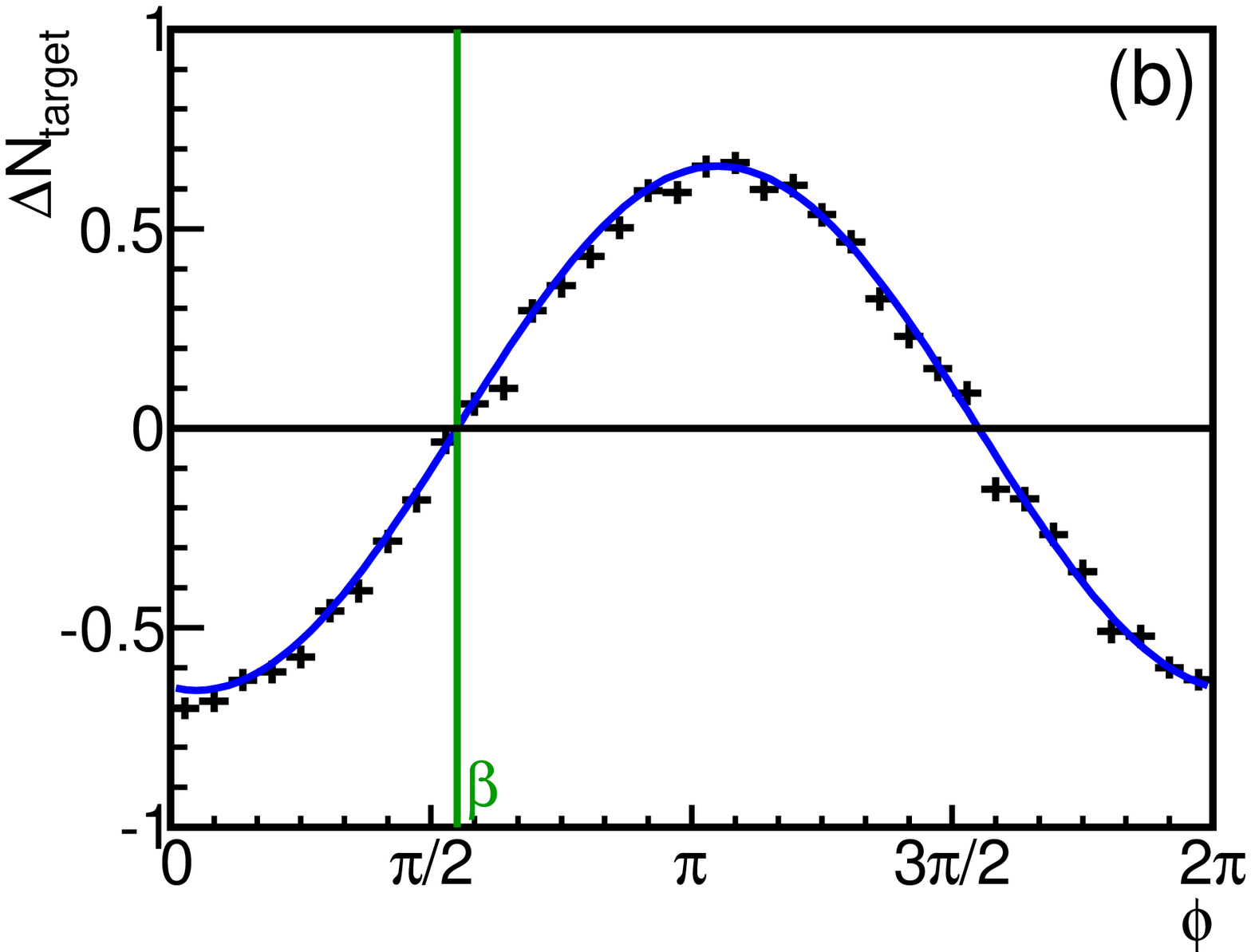}&
\hspace{-7mm}\includegraphics[width=0.35\linewidth]{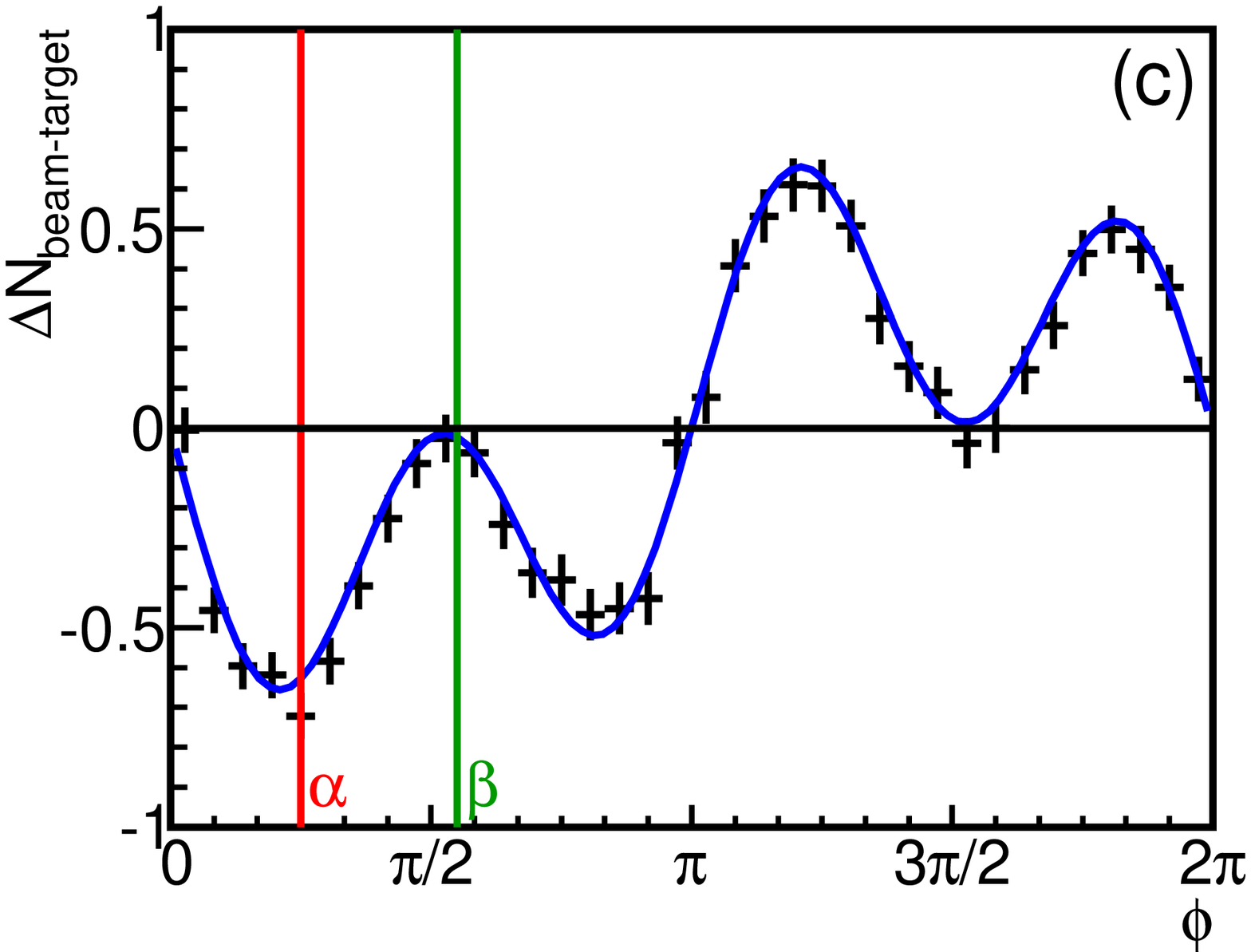}
\end{tabular}
\caption{\label{PTH-data}(Color online) Yield asymmetries as a function of
$\phi$ for the energy bin 1.46\,GeV$< W < 1.48$\,GeV. (a) $\Delta N_\text{beam}(\phi)$, (b) $\Delta N_\text{target}(\phi)$, (c) $\Delta N_\text{beam-target}(\phi)$ fitted by the function of Eq.~\eqref{pol-s}, Eq.~\eqref{pol-t}, and Eq.~\eqref{pol-ph}, respectively.
 }
\end{figure*}

In a first step, the ordinary beam asymmetry $\Sigma^\text{but}$ for each bin in energy and angle 
was determined:
\begin{equation}
\label{pol-s}
\Delta N_\text{beam}(\phi) = \frac{1}{p_{\gamma}} \cdot
\frac{N_\perp-N_\parallel}{N_\perp+N_\parallel} = \Sigma^\text{but} \cdot
\cos(2(\alpha-\phi)).
\end{equation}
A typical distribution is shown in Fig.~\ref{PTH-data}a. The
resulting beam asymmetries $\Sigma^\text{but}$ agree very well with previously
reported measurements \cite{Elsner:2008sn,Bartalini:2005wx,Sparks:2010vb} 
although the data sample contains in part reactions off nucleons of the 
C/O-nuclei. Results for $\Sigma^\text{but}$ are not shown here.

\begin{figure*}[t]
\begin{tabular}{ccc}
\hspace{-2mm}\includegraphics[width=0.35\linewidth]{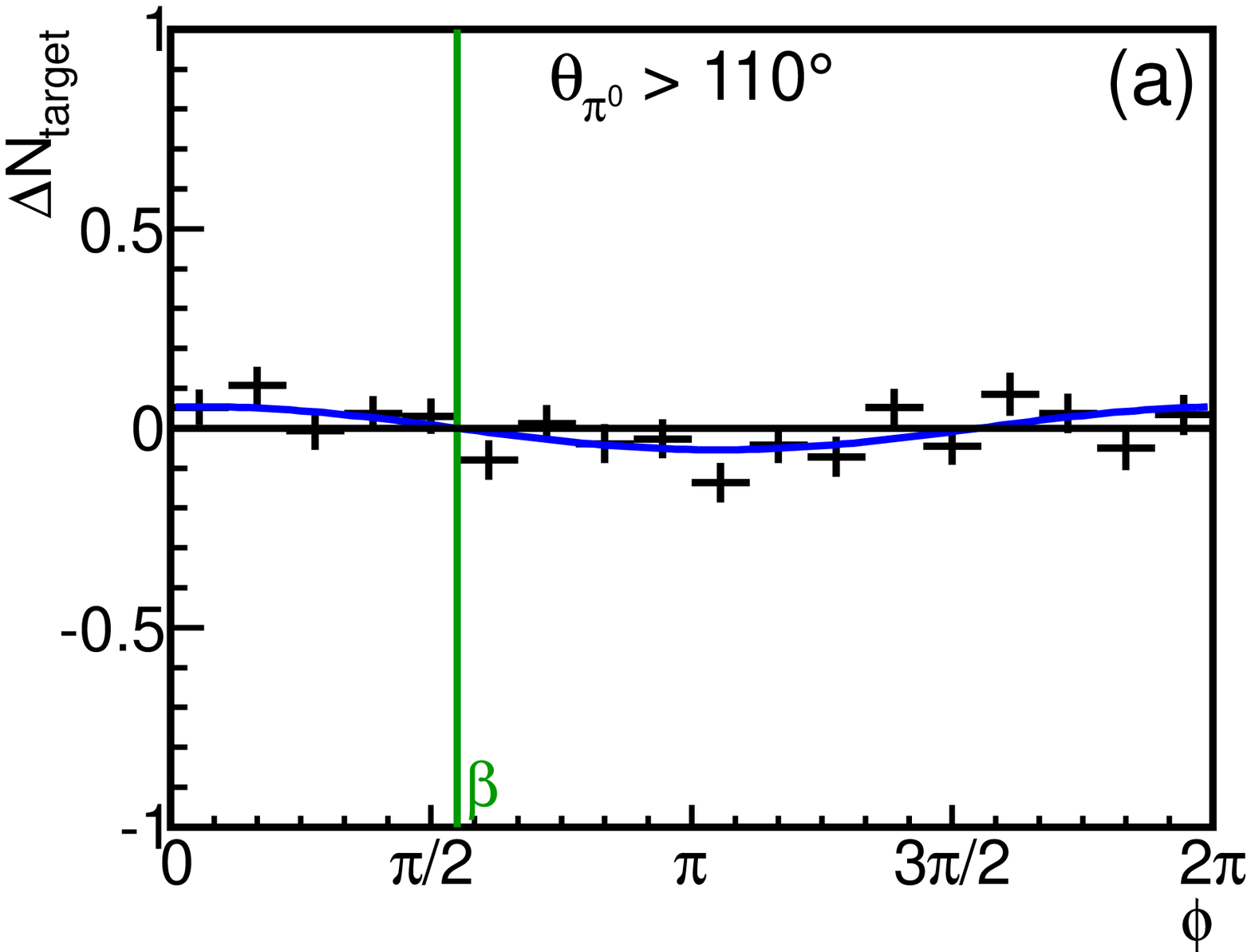}&
\hspace{-7mm}\includegraphics[width=0.35\linewidth]{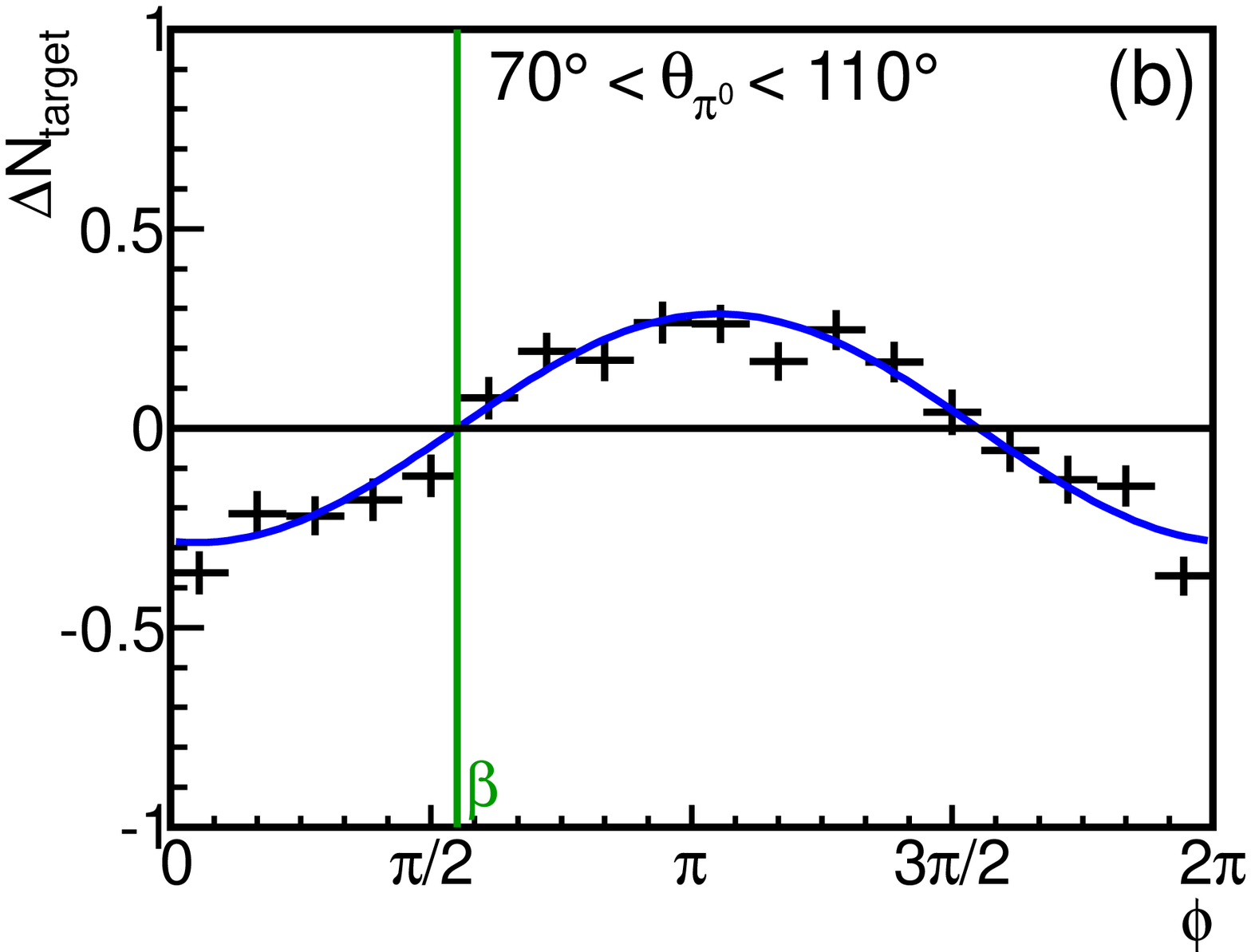}&
\hspace{-7mm}\includegraphics[width=0.35\linewidth]{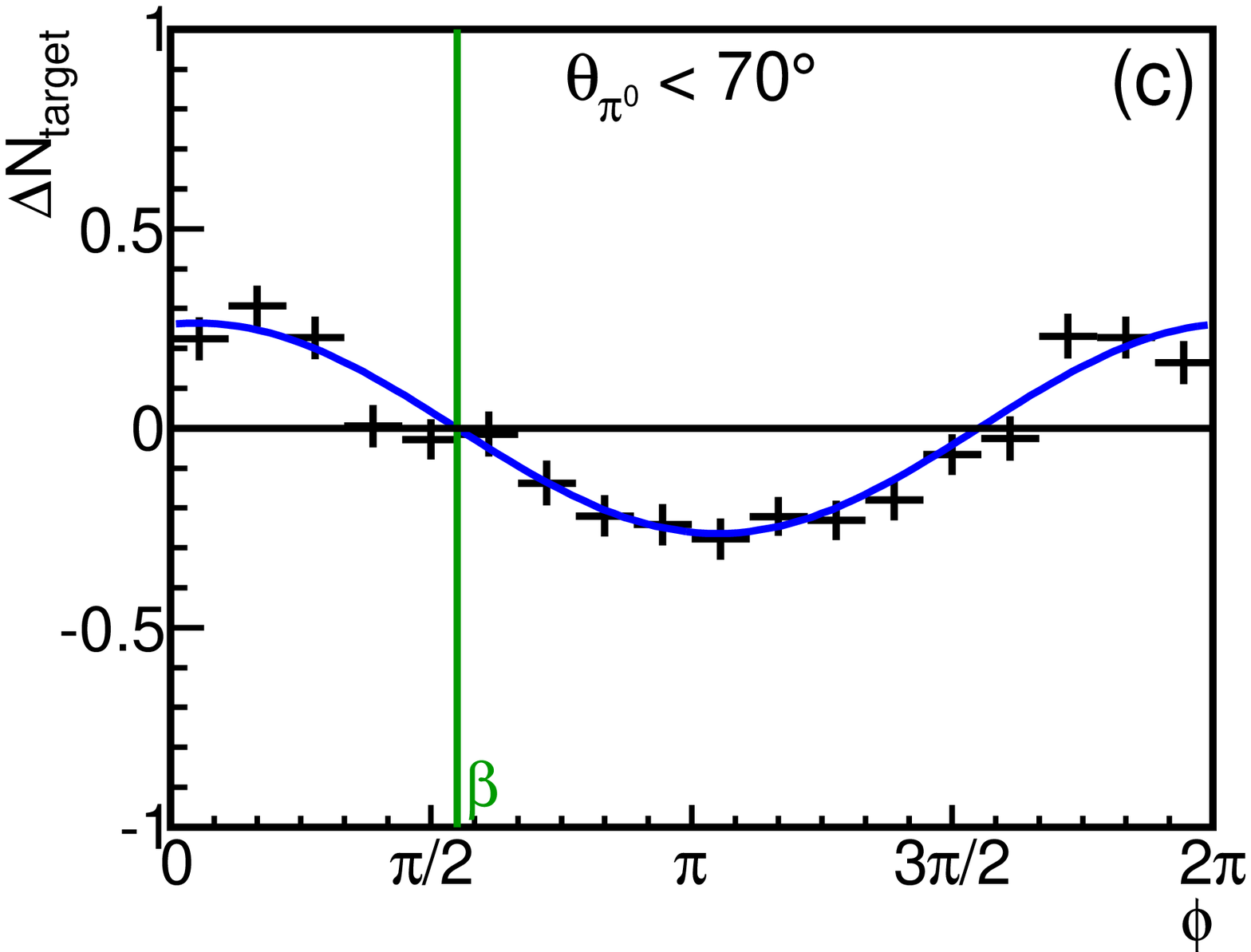}
\end{tabular}
\caption{\label{target}(Color online) Yield asymmetry $\Delta N_\text{target}$ as a function
of $\phi$ for three angular bins in $\theta$ in the $\gamma p$ invariant mass window 1.82\,GeV $ < W < 1.94$\,GeV.}
\end{figure*}

To determine $T$, the asymmetry of the data sets with respect to the
target polarization directions $\uparrow$ and $\downarrow$ was used,
resulting in
\begin{equation}
\label{pol-t}
\Delta N_\text{target}(\phi) = \frac{1}{d\,p_{\rm T}} \cdot
\frac{N_\uparrow-N_\downarrow}{N_\uparrow+N_\downarrow} = T \cdot
\sin(\beta-\phi)
\end{equation}
%with 
%\begin{equation}
%d = \dfrac{N_\text{butanol} - s \cdot N_\text{carbon}}{N_\text{butanol}}$.
%\end{equation}
with $d$ given by Eq.~\ref{dil_factor}.
The target asymmetry was determined by a fit to the $\Delta N_\text{target}(\phi)$
distributions as shown by the example in Fig.~\ref{PTH-data}b.
Figure~\ref{target} shows $\Delta N_\text{target}(\phi)$ for three 
different angular bins ($W$=1.82\,--1.94\,GeV).  
These distributions underline the strong dependence of $T$ on
the scattering angle $\theta$. 

$P$ and $H$ can be extracted from the data by considering the 
linear beam-polarization plane ($\parallel$ and $\perp$) in addition 
to the target polarization, leading to
\begin{align}
\label{pol-ph}
\Delta N_\text{beam-target}(\phi) ={}& \frac{1}{d\,p_{\gamma}\,p_{\rm T}} \cdot
\frac{(N_{\perp\uparrow}-N_{\perp\downarrow})-(N_{\parallel\uparrow}-
N_{\parallel\downarrow})}{(N_{\perp\uparrow}+N_{\perp\downarrow})+
(N_{\parallel\uparrow}+N_{\parallel\downarrow})} \nonumber\\
 ={}& P \sin(\beta-\phi)\cos(2(\alpha-\phi)) \nonumber \\
 & + H \cos(\beta-\phi)\sin(2(\alpha-\phi))
\end{align}
with average beam polarization $p_{\gamma}$ and
$\alpha=45^\circ$, the direction of the polarization plane in
the $\parallel$ setting. Again, the observables could easily be
determined by a fit to the $\Delta N_\text{beam-target}(\phi)$ distributions, see
Fig.~\ref{PTH-data}c.

\begin{figure*}[t]
\begin{tabular}{ccc}
\hspace{-2mm}\includegraphics[width=0.35\linewidth]{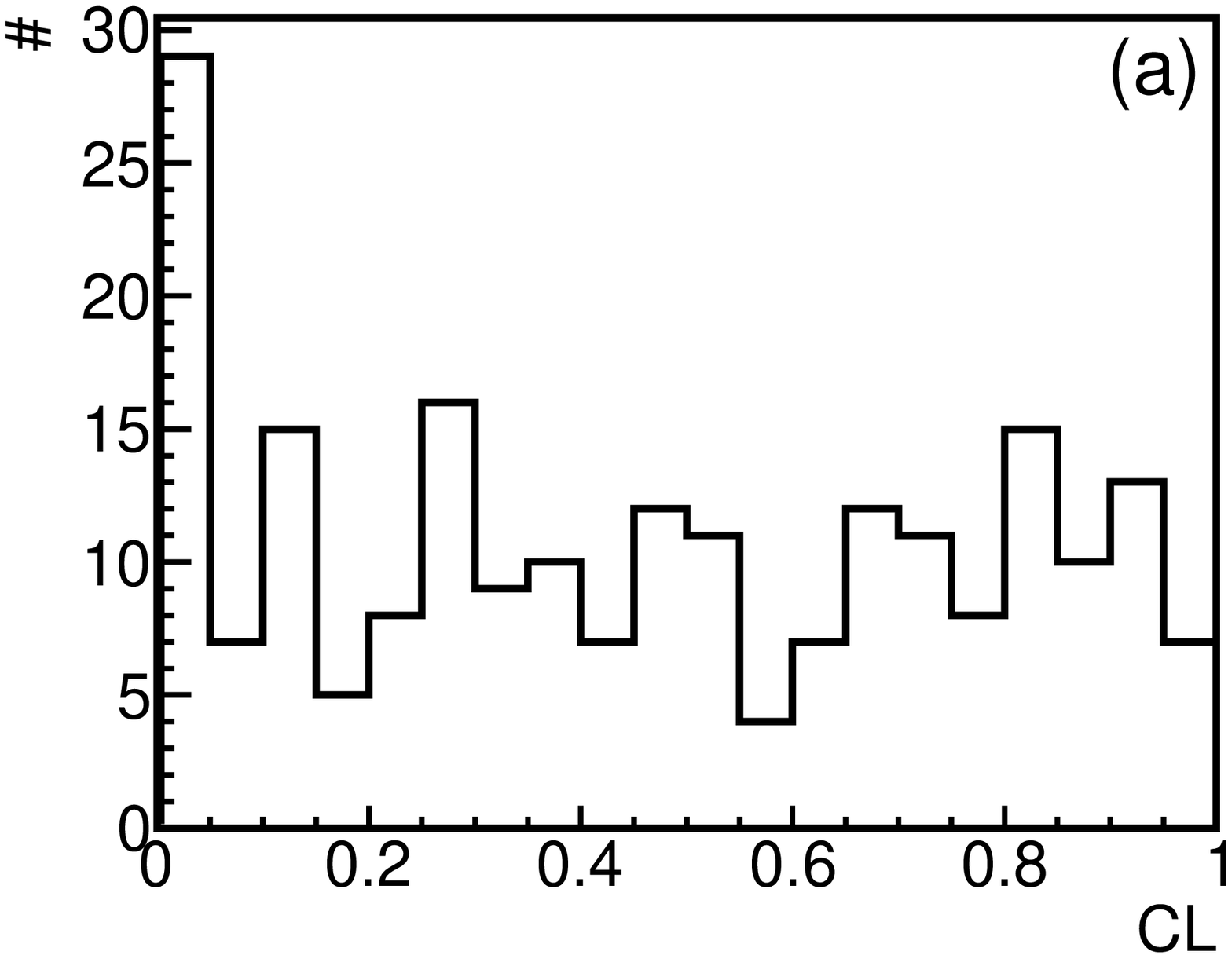}&
\hspace{-7mm}\includegraphics[width=0.35\linewidth]{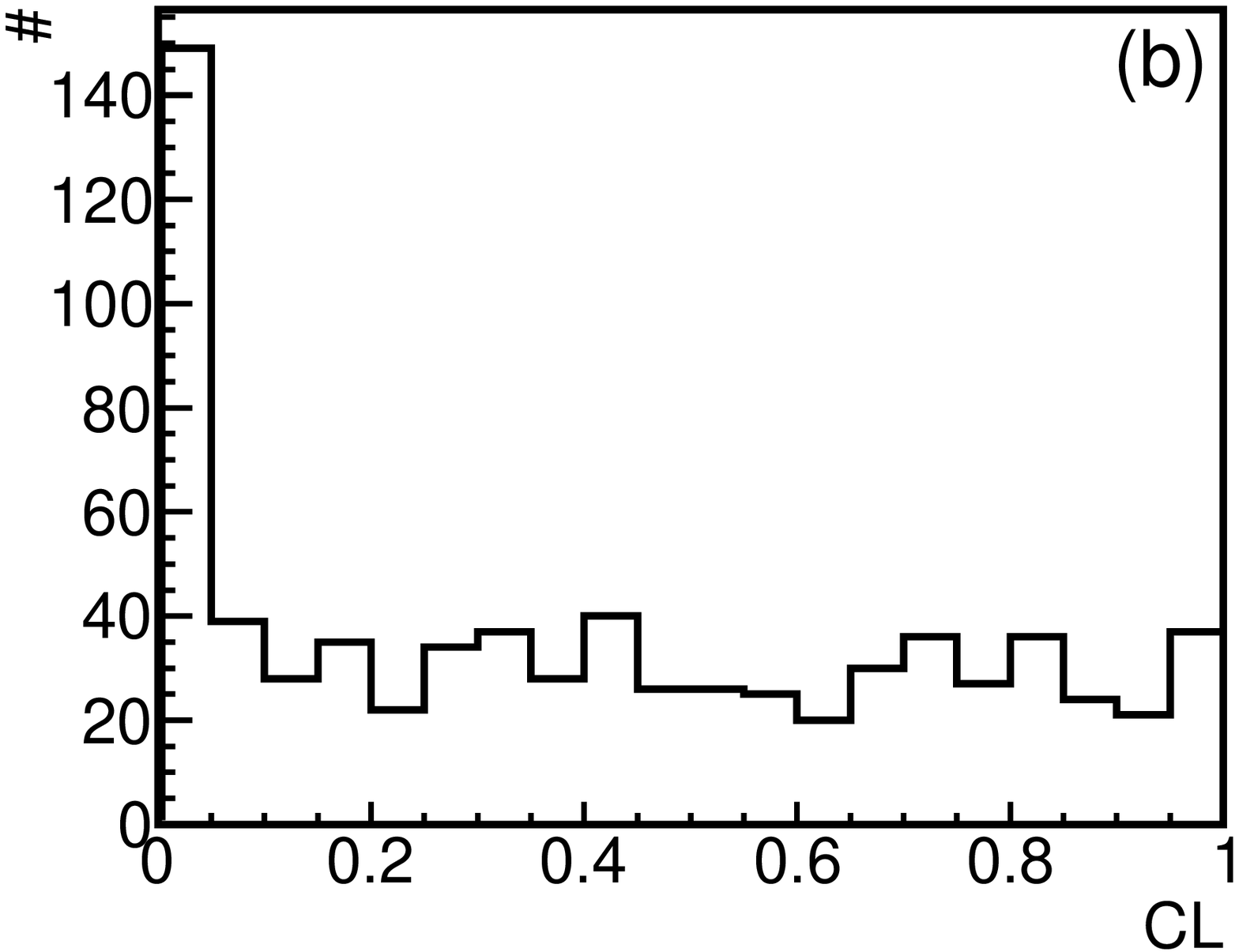}&
\hspace{-7mm}\includegraphics[width=0.35\linewidth]{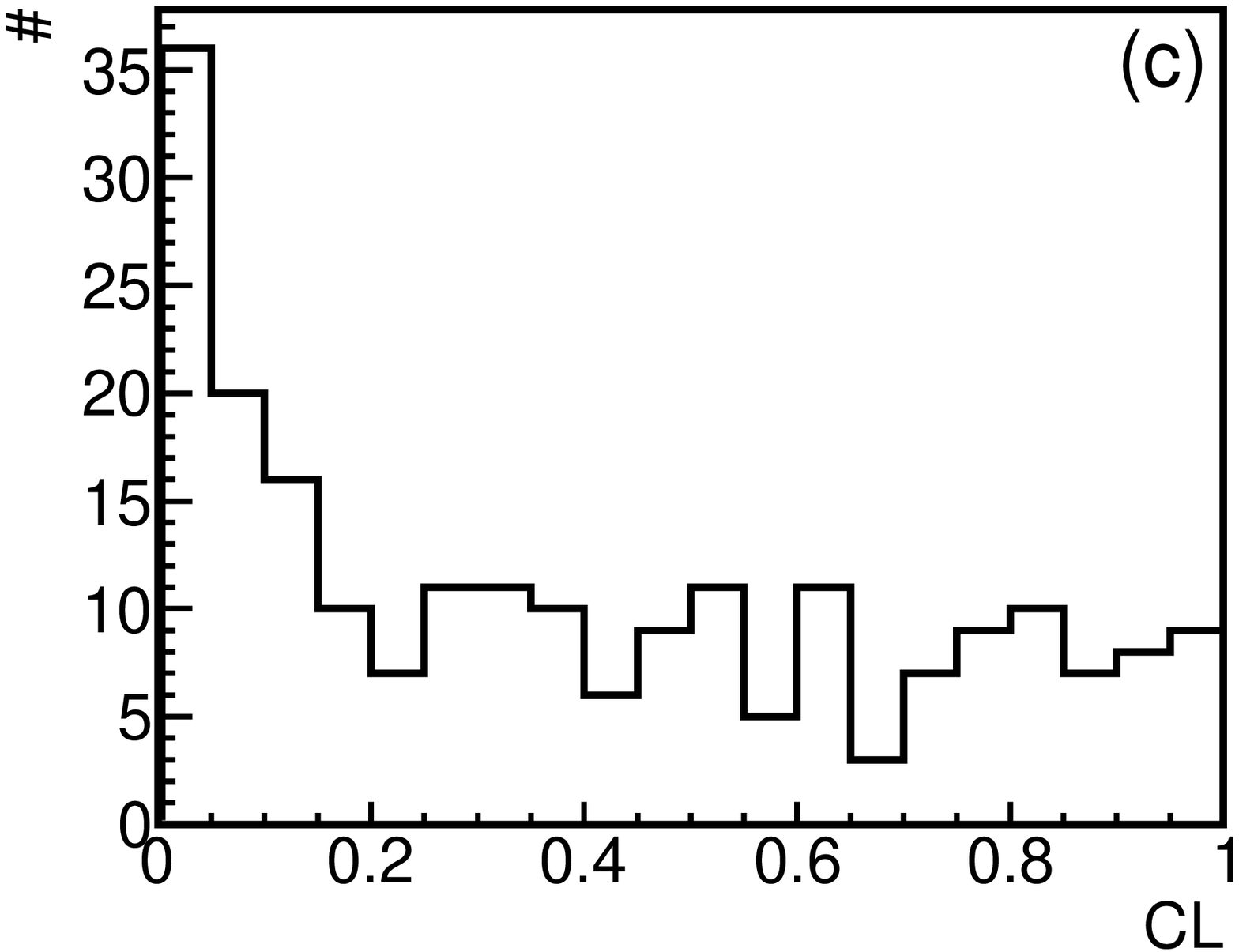}
\end{tabular}
\caption{\label{cl} Histograms of the confidence level of the fits to the (a) $\Delta N_\text{beam}(\phi)$, (b) $\Delta N_\text{target}(\phi)$, and (c) $\Delta N_\text{beam-target}(\phi)$ distributions. The increase of the CL in the first bin can be eliminated by requiring $\rm{CL} > 0.001$.}
\end{figure*}
The $\Delta N(\phi)$ distributions were fitted for each bin in energy $W$
and angle $\theta$. The resulting confidence level (CL) distribution
for the fits is flat with a distinct increase well below 0.1\%, as can be seen in Fig.~\ref{cl}. This increase 
was traced to fits of distributions with very low statistics 
(due to a low cross section or low acceptance). These fits were very sensitive 
to background fluctuations. These data were excluded from further analysis by
performing a CL-cut at 0.1\%.

\begin{figure*}[ht]
\begin{center}
\hspace*{-.01\textwidth}\includegraphics[width=0.95\textwidth]{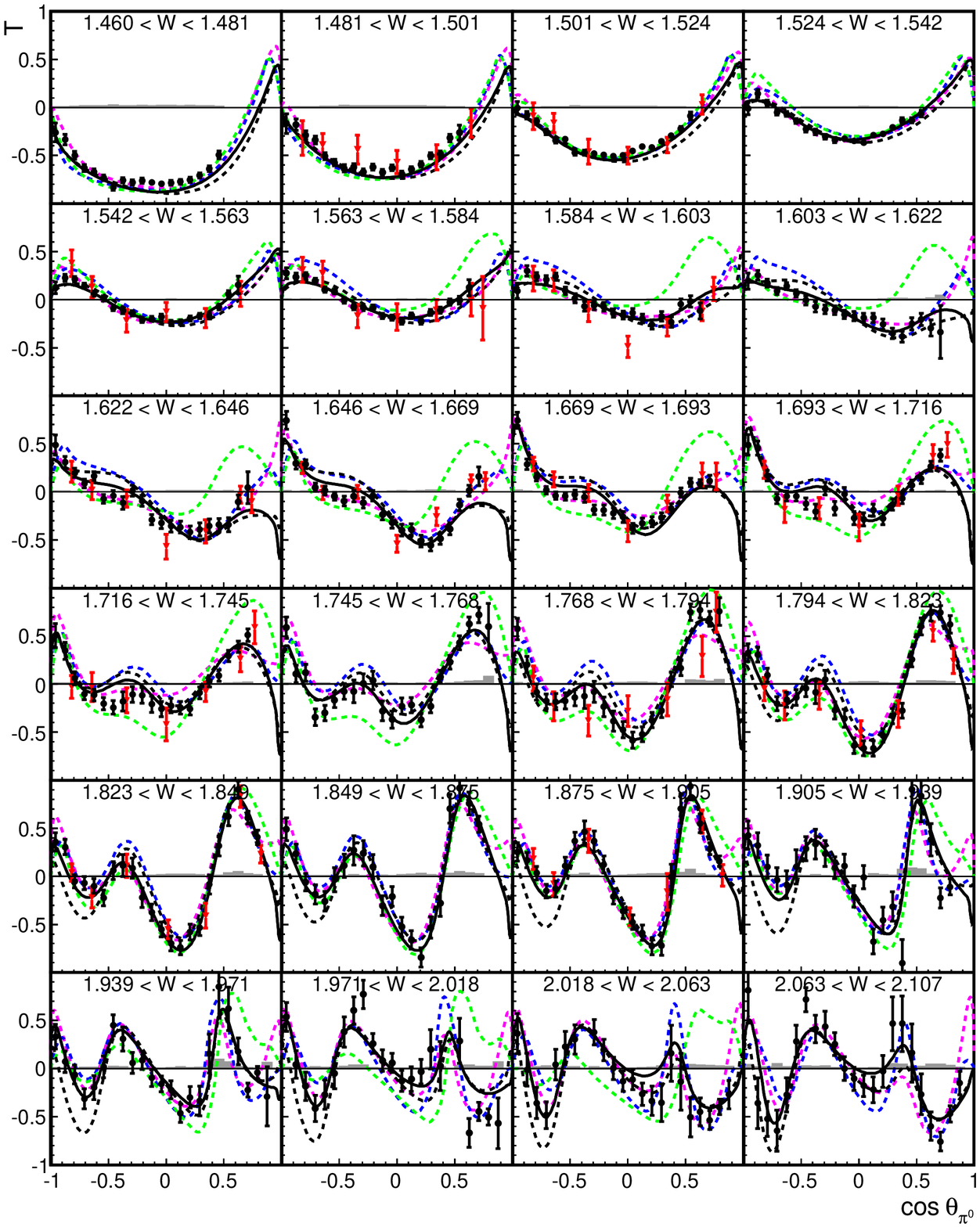}\\
\end{center}
\vspace*{-0.5cm}
\caption{\label{pic:t}(Color online) The polarization observable
$T$ as function of the $\gamma p$ invariant mass $W$ (in GeV)
and of the scattering angle $\cos\theta$. The systematic uncertainty is shown 
as dark gray band. An additional systematic uncertainty on the photon energy
(from $\sigma_{E_\gamma}^\text{sys}=$6.5~MeV at the lowest to 2.3~MeV at the 
highest energy bin) is not shown. 
The low energy data were presented in~\cite{jan_prl}. 
Earlier data (gray, (red, online)) are from \cite{Olddata:T}. The solid line represents our best fit BnGa2014.
The data are compared to predicitions (dashed curves) from BnGa2011-02 (black), MAID~\cite{Drechsel:2007if}
(light gray, (green, online)), SAID CM12~\cite{Workman:2012jf} (dark gray, (blue, online)), and Juelich2015~\cite{Ronchen:2015} (gray, (magenta, online)).}
\end{figure*}

\begin{figure*}[ht]
\begin{center}
\hspace*{-.01\textwidth}\includegraphics[width=0.95\textwidth,trim=0cm 0cm 0cm 15cm, clip]{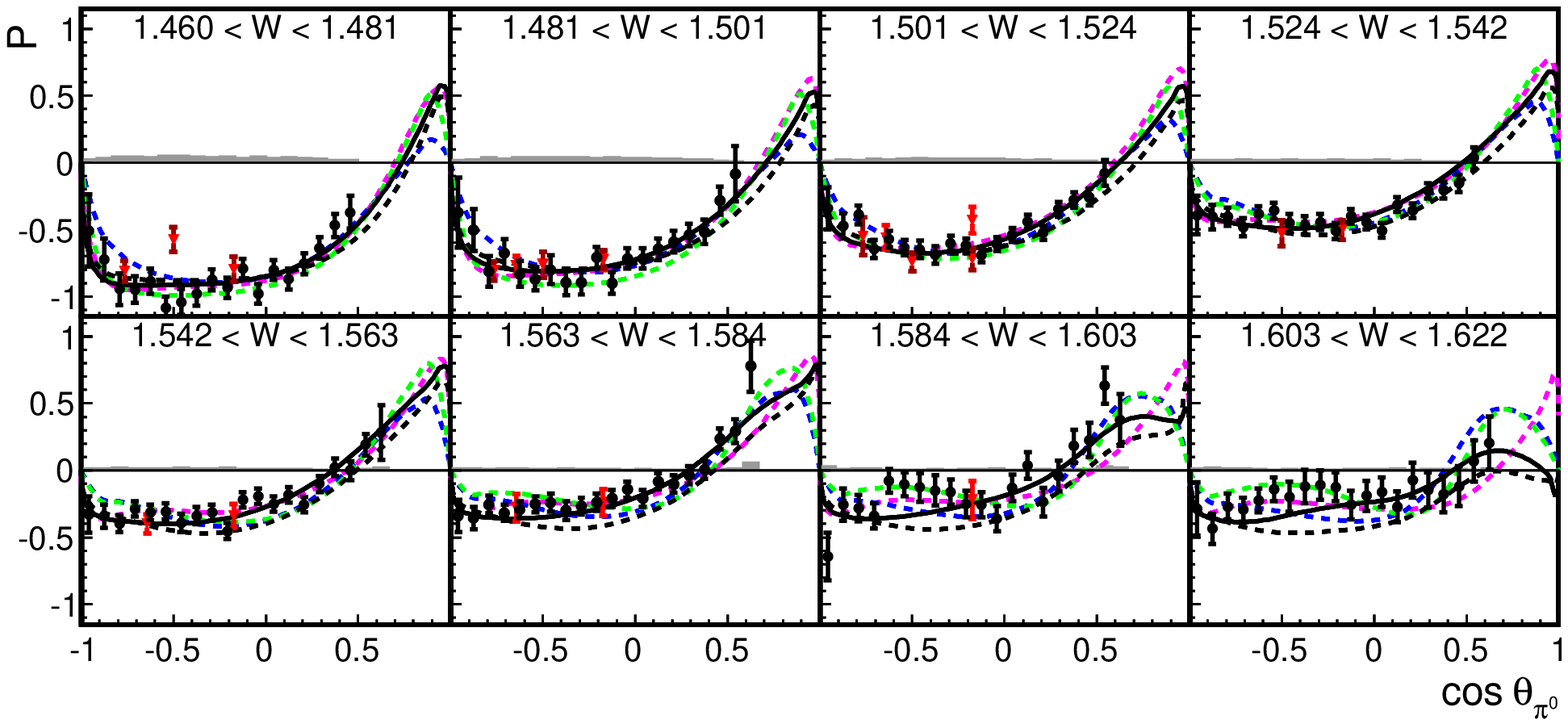}
\end{center}
\vspace*{-0.5cm}
\caption{\label{pic:p}(Color online) The polarization observable
$P$~\cite{jan_prl} as functions of the $\gamma p$ invariant mass $W$ (in GeV)
and of the scattering angle $\cos\theta$. The systematic uncertainty is shown 
as dark gray band.  An additional systematic uncertainty on the photon energy
(from $\sigma_{E_\gamma}^\text{sys}=$6.5~MeV at the lowest to 5.4~MeV at the 
highest energy bin) is not shown. Earlier data (gray, (red, online)) are
from \cite{Olddata:P}. The solid line represents our best fit BnGa2014.
The data are compared to predicitions (dashed curves) from BnGa2011-02 (black), MAID~\cite{Drechsel:2007if} 
(light gray, (green, online)), SAID CM12~\cite{Workman:2012jf} (dark gray, (blue, online)), and Juelich2015~\cite{Ronchen:2015} (gray, (magenta, online)).}
\end{figure*}

\begin{figure*}[ht]
\begin{center}
\hspace*{-.01\textwidth}\includegraphics[width=0.95\textwidth,trim=0cm 0cm 0cm 15cm, clip]{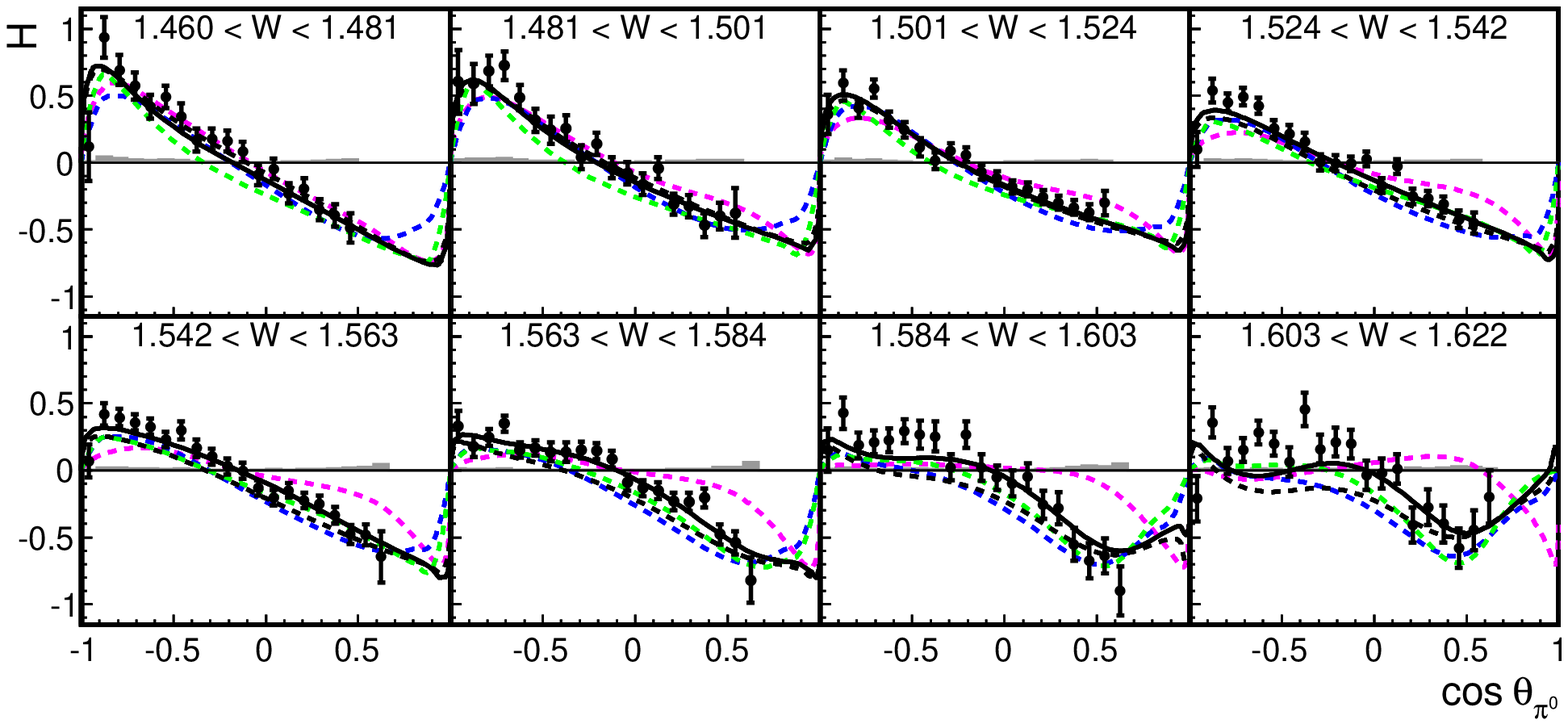}
\end{center}
\vspace*{-0.5cm}
\caption{\label{pic:h}(Color online) The polarization observable
$H$~\cite{jan_prl} as functions of the $\gamma p$ invariant mass $W$ (in GeV)
and of the scattering angle $\cos\theta$. The systematic uncertainty is shown 
as dark gray band. An additional systematic uncertainty on the photon energy
(from $\sigma_{E_\gamma}^\text{sys}=$6.5~MeV at the lowest to 5.4~MeV at the 
highest energy bin) is not shown. The solid line represents our best fit BnGa2014.
The data are compared to predicitions (dashed curves) from BnGa2011-02 (black), MAID~\cite{Drechsel:2007if}
(light gray, (green, online)), SAID CM12~\cite{Workman:2012jf} (dark gray, (blue, online)), and Juelich2015~\cite{Ronchen:2015} (gray, (magenta, online)).}
\end{figure*}

The data on $T$, $P$, and $H$ as functions of center-of-mass energy $W$ and
angle $\cos\theta$ are shown in Figs.~\ref{pic:t}--\ref{pic:h}. 
All observables exhibit a strong angular dependence which also changes
significantly with $W$. The systematic uncertainty shown 
includes the uncertainty in the degree of photon (4\%) and proton (2\%) polarizations, 
in the dilution factor (1\%--4\%), and an additional absolute uncertainty (0.01)
due to the remaining background contribution. For further details on the estimation of the systematic uncertainties see Refs. \cite{Elsner:2008sn,Dutz:1999,Hartmann:2015}.

The polarization observable $P$ describes the polarization of the
outgoing proton in the direction perpendicular to the scattering
plane. Here, it was determined indirectly from the correlation of
beam and target polarization. Thus, both observables, $P$ and $H$, were
measured only in the energy region in which the photon beam carried
a significant linear polarization; the results are hence restricted
to the 665 to 930\,MeV photon energy range. The target asymmetry $T$,
shown in Fig.~\ref{pic:t},
does not require polarized photons, and the data allowed us to
determine $T$ up to $E_\gamma = 1900$\,MeV. Above this energy, the count rates were
small, and we do not present those results here.

The new data agree well with previously reported measurements
of $P$ \cite{Olddata:P} and $T$ \cite{Olddata:T} but exceed the old data in precision and
coverage in angles and energy. For $H$ no earlier data exist in this energy range,
the older data \cite{Olddata:H} are limited to $E_\gamma > $  1300\,MeV.
Our data up to 930\,MeV were used as a basis for an energy independent PWA~\cite{jan_prl}. The high energy $T$-data are presented here for the first time. The new data sets have been included in the BnGa-PWA as discussed in the following.

\section{Partial wave analysis}

\begin{figure*}[ht]
\begin{center}\vspace*{-1.5cm}
\hspace{-2mm}\includegraphics[width=.85\textwidth]{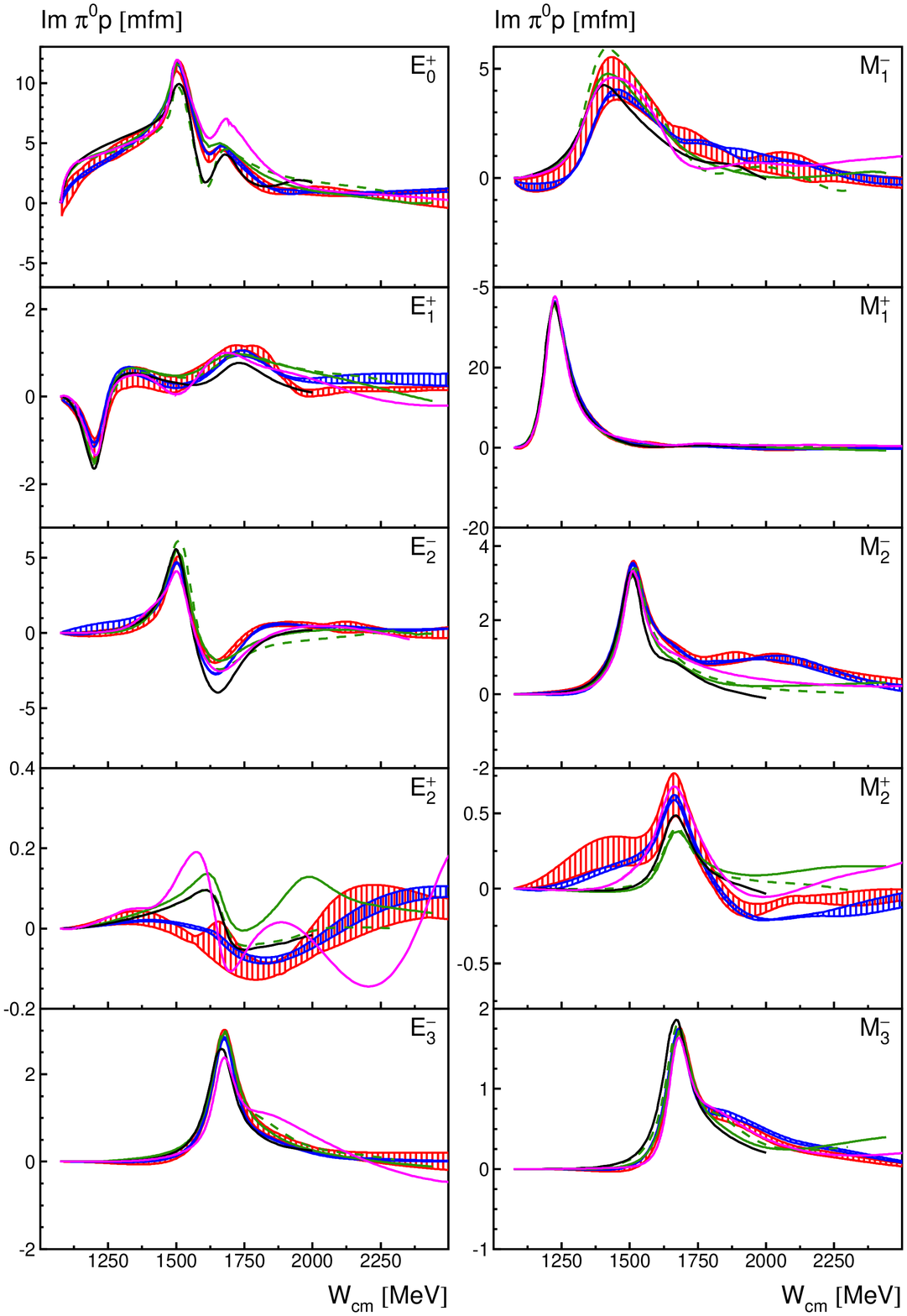}
\end{center}\vspace*{-.5cm}
\caption{\label{piNa}(Color online) Multipole decomposition of the $\gamma p\to
p\pi^0$ transition amplitudes, imaginary part. The light (red)
shaded areas give the range from a variety of different fits derived
from solution BnGa2011-01 and BnGa2011-02~\cite{Anisovich:2011fc}. The dark (blue) shaded
area represents the range of solutions when the new data are
included in the fit. The black curves represent the MAID fit
\cite{Drechsel:2007if}, the light (green) solid curves SAID-CM12~\cite{Workman:2012jf},
the light (green) dashed curves SAID-SN11~\cite{Workman:2011vb}, and the magenta curve the
Juelich2015~\cite{Ronchen:2015} solution.  For the BnGa-multipoles an error band (see text) 
has been determined. Such an error band is presently not provided by the other analyses.}
\end{figure*}
\begin{figure*}[ht]
\begin{center}\vspace*{-1.5cm}
\hspace{-2mm}\includegraphics[width=.85\textwidth]{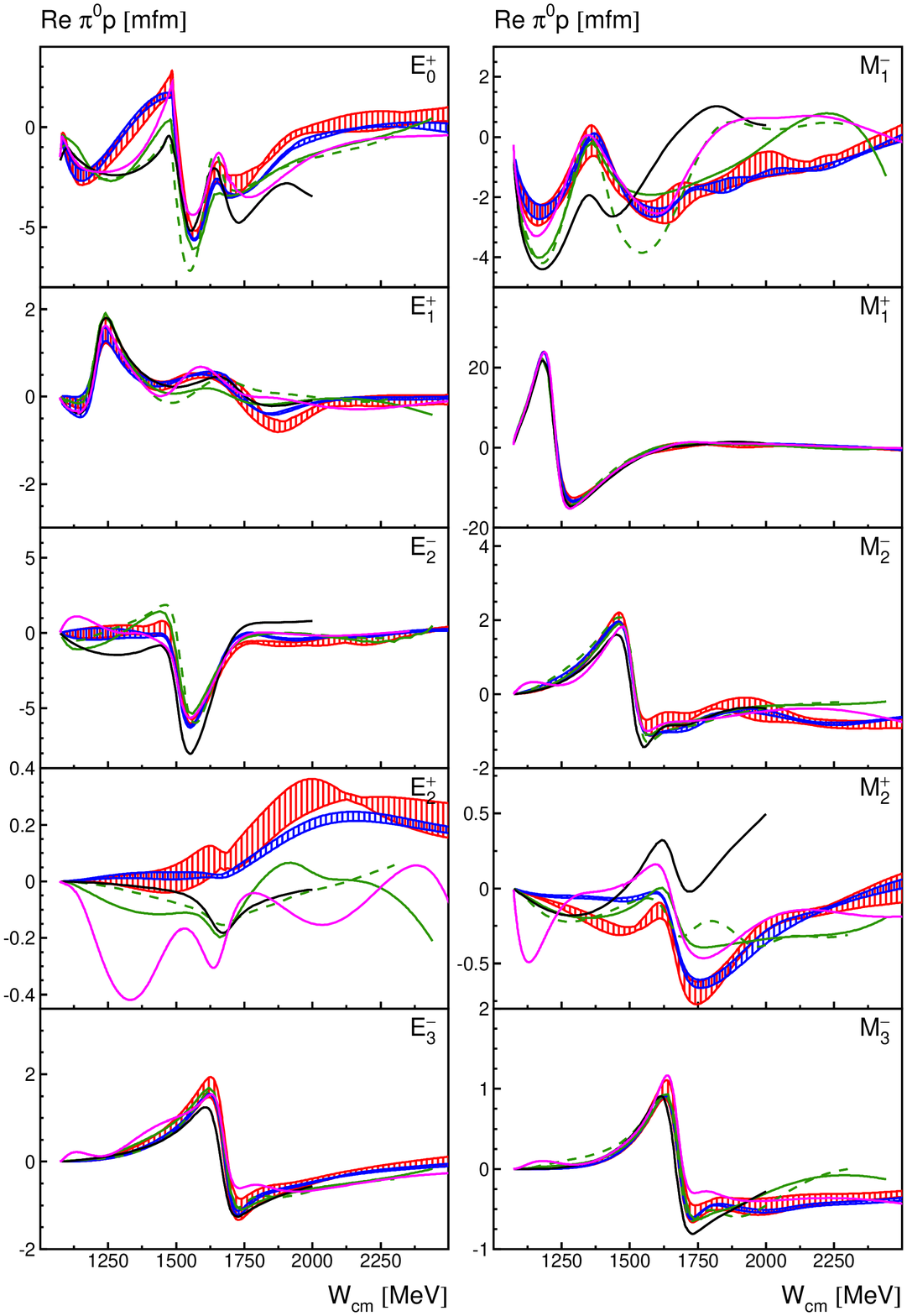}
\end{center}
\vspace*{-.5cm}\caption{\label{piNb}(Color online) Multipole decomposition of the $\gamma p\to
p\pi^0$ transition amplitudes, real part. The light (red) shaded
areas give the range from a variety of different fits derived from
solution BnGa2011-01 and BnGa2011-02~\cite{Anisovich:2011fc}. The dark (blue) shaded area
represents the range of solutions when the new data are included in
the fit. The black curves represent the MAID fit
\cite{Drechsel:2007if}, the light (green) solid curves SAID-CM12~\cite{Workman:2012jf},
the light (green) dashed curves SAID-SN11~\cite{Workman:2011vb}, and the magenta curve the
Juelich2015~\cite{Ronchen:2015} solution. For the BnGa-multipoles an error band (see text) 
has been determined. Such an error band is presently not provided by other analyses.}
\end{figure*}

%\clearpage
The data were fitted within the BnGa multi-channel partial wave
analysis. Compared to our solutions BnGa2011-01 and BnGa2011-02,
further data were included~\cite{Thiel:2012yj,Gottschall:2014,data_added_for_BnGa2014}.
%, in particular on hyperon production~\cite{Anisovich:2013vpa}.
Relevant for the $\gamma p\to p\pi^0$ multipoles are the new data 
on $T$, $P$, and $H$ as well as our recently published data on 
$G$ \cite{Thiel:2012yj} and on $E$ \cite{Gottschall:2014}.
Figures~\ref{piNa} and \ref{piNb} compare the newly determined 
multipoles with those of BnGa2011-01 and BnGa2011-02. The error bands for the
BnGa2011 solutions were derived from the ($1\sigma$) spread of
12 different solutions with different assumptions on the ingredients: the
number of poles in the $J^P=3/2^+$ wave was 3 or 4, in the
$J^P=5/2^+$ wave 2 or 3, the $N(1700)3/2^-$ width of the pole
converged to a wide ($\sim$600\,MeV) or a narrow ($\sim$250\,MeV)
value, a K-matrix formalism was used or, alternatively, an $N/D$-parametrization~\cite{N_over_D}. 
%{\color{red} \bf Can we decide by our new fits including the polarisation 
%data on the the number of poles for one of these waves ??? ... if yes 
%this would be a strong statement .... could we maybe even show how the 
%$\chi^2$-minimum gets more convincing with the new data included compared 
%to before? }
%
%
The $\chi^2$ ranged from its minimum $\chi^2_{\rm min}$ to
$\chi^2_{\rm min}+800$. (Note that the absolute $\chi^2$ value is
meaningless since part of the data are multiparticle final states
and fitted in an event-based likelihood fit. The log likelihood
value is then converted into a pseudo-$\chi^2$~\cite{Anisovich:2011fc}.)

In the new fits, we started from the same solutions and re-optimized
all parameters. All fits converged, but 6 fits resulted in a $\chi^2$ larger
than the new $\chi^2$-minimum ($\chi^2_{\rm new\,min}$) 
by 1000 units or more. These fits, mostly
those with only 3 poles in the $J^P=3/2^+$ wave, were then removed
from the error analysis. The resulting error bands for 
all remaining solutions within $\chi^2_{\rm new\,min}+800$ 
are also shown in Figs.~\ref{piNa} and ~\ref{piNb}. 
The $\chi^2$-range for the new solutions was chosen consistently with the 
BnGa2011 solutions for which the multipoles have first been shown in~\cite{Anisovich:2012ct}.
%{\color{red} \bf (We do not eliminate one of the old solutions completely??? ... 
%I assume not, right ?)}

The new error bands are significantly smaller than the previous ones. 
Averaged over all multipoles and energies, the errors are reduced by
a factor of 2.25. {Examples for multipoles which are 
substantially better defined are the  $M_1^-$ multipole 
to which the Roper resonance and $N(1710)1/2^-$ contribute and 
the $M_2^+$ multipole to which the $J^P=5/2^-$-resonances contribute.} 
In most cases, the old error bars cover the range of new
solutions, the solutions are at least compatible with each other at
the $2\sigma$ level over the full mass range. 

There are few changes to the multipoles only: The $E_1^+$ multipole 
leading to the excitation of $N$ and
$\Delta$ resonances with $J^P=3/2^+$ has kept its structure but its
imaginary part has increased in strength in the fourth resonance
region. {Similar changes are also visible in the real part.} 
The real part of the $M_2^+$ multipole ($J^P=5/2^-$) has
become smaller in absolute value in the $W=1500\,MeV$ region, {while in the 
imaginary part changes are observed in the high mass region around $W=2100\,MeV$}. 
In all cases, where 
discrepancies with the MAID and SAID solutions were observed, these
discrepancies remain while the consistency between BnGa2011 and
BnGa2014 is rather good. The multipoles from 
Juelich2015~\cite{Ronchen:2015} show also significant differences compared 
to BnGa2014.
%
%If the new polarization data is also included in the different other 
%analysis one might expect also a step forward toward a commonly 
%accepted answer how the different p$\pi^0$ multipoles look like.  
%

Summarizing, we have reported a simultaneous measurement of the polarization
observables $T$, $P$, and $H$. With the data presented here an 
additional step toward a complete 
experiment in $\pi^0$-photoproduction off the proton has been made. 
The data provide a more precise
determination of the photoproduction multipoles governing the
photoproduction of single neutral pions off protons.

\section*{Acknowledgement} We thank the technical staff of ELSA, the polarized 
target group, and the par\-ti\-ci\-pating institutions for their invaluable
contributions to the success of the experiment. We acknowledge
support from the \textit{Deutsche Forschungsgemeinschaft} (SFB/TR16)
and \textit{Schweizerischer Nationalfonds}.

\end{document}